\documentclass[preprint,3p,12pt]{elsarticle}
\usepackage{mathrsfs}
\usepackage{amsmath}
\usepackage{stmaryrd}
\usepackage{bbding}
\usepackage{dcolumn}
\usepackage{graphicx}
\usepackage{amsfonts}
\usepackage{amssymb}
\usepackage{psfrag}
\usepackage{wrapfig}
\usepackage{subfigure}
\usepackage{makeidx}
\usepackage{bm}
\usepackage{epsf}
\usepackage{epsfig}
\usepackage{setspace}
\usepackage{graphicx}
\usepackage{epstopdf}
\usepackage{psfrag}
\usepackage{subfigure}
\usepackage{color}
\usepackage{comment}
\usepackage{caption}
\renewcommand{\vec}[1]{{\boldsymbol{#1}}}

\begin{document}

\title{Wave-Particle Based Multiscale Modeling and Simulation of Non-equilibrium Turbulent Flows }

\author[HKUST1]{Xiaojian Yang}
\ead{xyangbm@connect.ust.hk}

\author[HKUST1,HKUST2,HKUST3]{Kun Xu\corref{cor1}}
\ead{makxu@ust.hk}

\address[HKUST1]{Department of Mathematics, Hong Kong University of Science and Technology, Clear Water Bay, Kowloon, Hong Kong}
\address[HKUST2]{Department of Mechanical and Aerospace Engineering, Hong Kong University of Science and Technology, Clear Water Bay, Kowloon, Hong Kong}
\address[HKUST3]{Shenzhen Research Institute, Hong Kong University of Science and Technology, Shenzhen, China}
\cortext[cor1]{Corresponding author}

\begin{abstract}

This paper presents a novel methodology for the direct numerical modeling and simulation of turbulent flows.
The kinetic model equation is firstly extended to turbulent flow with the account of coupled evolution of kinetic, thermal, and turbulent energy.
Based on the kinetic model, a unified framework for the laminar and turbulent flow is constructed through wave-particle decomposition, following the coupled dynamic evolution of wave and particles. With the consideration of multiscale flow structure of the turbulent flow and the numerical observation scale,
central to this methodology is the use of numerical cell size, time step, and local resolved flow strain rate tensor to determine the degrees of freedom required to represent turbulence within each control volume. The modeling is based on the assumption that the turbulence emergence is attributed to the breakdown of continuously connected fluid elements under the cell resolution. Then, the non-equilibrium transport of discrete fluid elements carrying the turbulent kinetic energy is constructed through the particle movement.
The model leverages a hybrid wave-particle representation, where wave dynamics governed by the Navier-Stokes equations provide a background resolved flow structure, while particle transport is driven by the unresolved turbulent dynamics. Particle non-equilibrium trajectory crossing, collision, and interaction with the background wave, distinguish the current model from conventional RANS and LES methodologies for the unresolved turbulent flow simulation.
The large eddy viscosity coefficient and the cell resolution determine the particle number and quantify the non-equilibrium transport of turbulent eddies.
Instead of the dissipation model constructed in the mixing length theory, the present model presents a upgrading non-equilibrium transport model with particle penetration and collision. The transition between laminar and turbulent states is determined by the particle density in the wave-particle decomposition inside each cell.
Notably, the absence of particles reduces the current unified gas-kinetic wave-particle  for turbulent simulation (WPTS) method to the gas-kinetic scheme (GKS), ensuring seamless recovery of laminar Navier-Stokes solutions. Emphasizing genuine non-equilibrium particle transport, the proposed multiscale method demonstrates enhanced accuracy for capturing the turbulent flow, where the Reynolds stress can be directly obtained from the flow field in the study of compressible mixing layer.
This work offers a versatile tool for turbulence research with potential applications in aerospace, energy systems, and environmental fluid dynamics.

\end{abstract}


\maketitle

\section{Introduction}

The study of turbulent flows represents one of the most challenging and fundamental problems in fluid mechanics. Beginning with the Navier-Stokes equations in the 19th century, researchers sought solutions that could capture the complex, chaotic behavior of turbulent flows. The deterministic approach relies on solving the full Navier-Stokes equations through Direct Numerical Simulation (DNS), attempting to resolve all spatial and temporal scales of turbulent motion. However, this method becomes computationally prohibitive at high Reynolds numbers due to the vast range of scales involved.
This limitation led to the development of Reynolds-averaged Navier-Stokes (RANS) methods in the early 20th century, which decompose flow variables into mean and fluctuating components \cite{lesieur2005large,wilcox1998turbulence,sagaut2013multiscale}. While more computationally tractable, RANS approaches introduce the closure problem, requiring various turbulence models to approximate unknown Reynolds stress terms. Common models like $k-\epsilon$ and Reynolds stress models provide practical engineering solutions but sacrifice detailed information about turbulent structures. As computational power increased, Large Eddy Simulation (LES) emerged as an intermediate approach between DNS and RANS. LES explicitly resolves larger energy-containing eddies while modeling smaller, more universal scales through subgrid-scale models. This method better captures transient phenomena and coherent structures but still requires significant computational resources \cite{nieuwstadt2016turbulence}.

Recognition of turbulence's inherently random nature gradually shifted focus toward statistical descriptions \cite{tennekes1972first,frisch1995turbulence,leschziner2015statistical}. This paradigm change acknowledged that while individual realizations of turbulent flows might be unpredictable, their statistical properties often show remarkable regularity. This led to the development of moment closure methods and the introduction of probability concepts in turbulence modeling. The evolution culminated in the development of probability density function (PDF) methods, which represent a more complete statistical description of turbulent flows. PDF approaches directly solve for the joint probability distribution of velocity and scalar quantities, naturally incorporating chemical reactions without requiring additional closure assumptions \cite{pope2001turbulent}. These methods, available in both Eulerian and Lagrangian frameworks, provide a powerful tool for studying complex turbulent reacting flows.

This evolution from deterministic to stochastic approaches reflects not only advancing computational capabilities but also a fundamental shift in our understanding and modeling of turbulent phenomena. Contemporary turbulence research continues to progress along multiple paths, synthesizing insights from deterministic, statistical, and probabilistic approaches to develop more accurate and efficient prediction methods {\cite{guo2021dual}.

This paper presents a direct modeling methodology for turbulent flow simulation, focusing on macroscopic turbulent flow. The approach draws from non-equilibrium multiscale modeling concepts developed for rarefied flow. It is believed that the non-equilibrium transport characteristics in rarefied and turbulent flows share notable similarities. In rarefied flow, individual particle transport and insufficient collisions maintain the system in a non-equilibrium state, while the multiscale method is able to connect the kinetic-scale particle transport to the equilibrium wave propagation in unified framework. Here the similar methodology will be developed to connect the laminar and turbulent flows seamlessly \cite{xu2021unified}.

In order to model and simulate turbulent flow, a kinetic model equation will be extended from the rarefied flow to the turbulent one. Then, the direct modelling methodology will be developed to solve the kinetic equation in different scales simultaneously. The simulation method will borrow the ideas from the non-equilibrium multiscale method for the rarefied flow, especially the canonical form of the unified gas-kinetic wave-particle (UGKWP) method \cite{WP-first-liu2020unified}.
For the turbulent flow, the main picture is that at high Reynolds number the large strain rate between fluid elements will gradually break down their connection.
The breakdown of the element connection is possibly related to the formation of singularity in the Navier-Stokes solution \cite{dou2022origin}.
As a result, the emerging particle like bulk of fluid elements will transport freely under a continuous wave background. The dynamic behaviours of the particle part of the turbulent flow will be similar to the Prandtl’s mixing length theory, but the nonequilibrium transport will be fully followed in the current model, instead of the modification of dissipative coefficient in the original Prandtl’s model. The advances of the computer power make it possible to directly construct the non-equilibrium model and simulate the turbulent flow.

The target of this paper is to decompose the turbulent description into particle and wave and model their coupled dynamic evolution. As a result, there is no clear distinction between laminar and turbulent flow. The emerging and disappear of particles determine the transition among the laminar and turbulent flow, and the particle number determines the local intensity of turbulence. In other words, the current model will unify the deterministic and stochastic models for the turbulent to capture the well-organized background flow and the local stochastic behaviour. The development of the current methodology for turbulent flow modelling and simulation are solely from the three decades accumulation on the development of gas-kinetic scheme (GKS) for the Navier-Stokes solution \cite{GKS-2001}, and the unified gas-kinetic scheme (UGKS) and unified gas-kinetic wave-particle (UGKWP) method for the rarefied and continuum flows \cite{UGKS-xu2010unified,wei2023adaptive}. The dynamic modelling is to use the spatial cell resolution, local fluid strain rate, and the numerical time step to determine the wave-particle decompositions and subsequently develop their dynamic evolution in the discretized space. With the absence of the particle in the well-resolved flow region, the current modelling will recover the GKS for the laminar Navier-Stokes solutions automatically. In the turbulent flow region, the particles will provide the non-equilibrium transport and recover the observation of the turbulent flow, such as the capturing of laminar-turbulence transition and direct evaluation of  nonequilibrium Reynolds stress. The coherent structure will be observed from the background wave distribution and the forward and inverse cascades are associated with replenishing and annihilating of  particles.

This paper will be arranged in the following. Section 2 is about the introduction of the kinetic model for turbulent flow.
Section 3 introduces the methodology in determining the parameters in kinetic model and constructing the corresponding numerical scheme. The current wave-particle turbulent simulation (WPTS) method will recover the gas-kinetic scheme (GKS) for NS solution in the laminar flow limit.
Section 4 shows the numerical test.
Then the last section is the conclusion.

\section{Turbulent multiscale non-equilibrium modeling}	

The current development of modelling of turbulent flow is coming from the accumulating investigation in the direct modelling for the computational fluid dynamics (CFD). Recent decades have witnessed the development of a unified CFD algorithm spanning both rarefied and continuum flow simulations. The gas-kinetic scheme (GKS) was firstly developed for the laminar Navier-Stokes solutions. In order to extend it to rarefied flow, the non-equilibrium gas distribution function is tracked in unified gas-kinetic scheme (UGKS), rather than the reconstructed Chapman-Enskog expansion. The UGKS simultaneously updates macroscopic flow variables and gas distribution functions and its solution depends on the cell’s Knudsen number, which is the ratio of particle mean free path over the cell size. In other words, once the cell size is in the kinetic scale of the particle mean free path, the Boltzmann solution will be captured. When the cell size is in the hydrodynamic scale, the Navier-Stokes solution is obtained. As a discrete velocity method (DVM), the UGKS discretizes particle velocity space using grid points. The distinguishable feature of UGKS from traditional DVM method is that dynamic effect of particle collision is explicitly included in the particle transport process in the evaluation of cell interface numerical fluxes. The UGKS fully integrates macroscopic flow variable and gas distribution function updates, utilizing integral solutions of the kinetic model equation in encompassing both particle free transport and hydrodynamic wave propagation \cite{xu2021unified}.

The UGKS implements direct modeling using cell Knudsen number to determine flow evolution, with mesh size and time step serving as fluid dynamic modeling scales. This versatility enables handling diverse flow physics across regions, from compressible Navier-Stokes solutions to free molecular flow. For high-speed flows requiring numerous grid points to capture gas distribution variations across extreme temperature ranges, computational efficiency concerns led to the development of the unified gas-kinetic particle (UGKP) method and its enhancement, the unified gas-kinetic wave-particle (UGKWP) method \cite{WP-first-liu2020unified,WP-second-zhu-unstructured-mesh-zhu2019unified}. The UGKWP optimizes efficiency through selective particle tracking and analytical treatment of collisional transport as waves. The UGKWP's effectiveness stems from its adaptive nature, varying particle-wave coupling with local Knudsen numbers. It achieves seamless transition between flow regimes while maintaining computational efficiency comparable to specialized methods. This comprehensive framework accommodates complex flows by incorporating both non-equilibrium transport and traditional diffusive processes. For the non-equilibrium and multiscale turbulent flow, we believe that the same methodology of UGKWP can be extended here as well in the construction of a new turbulent modelling with the consideration of both deterministic and stochastic feature of turbulent flow.

The continuum flow dynamic equations are derived from the model of continuous connected fluid elements. The isolated fluid element keeps the same mass and transport with its neighbouring elements. As indicated in the Lagrangian formulation, the neighbouring fluid elements are always keeping the same topological connection. However, in our new turbulent modelling, we believe that the connection between fluid elements can break down and the long distance transport of the element provides the dynamics of  the high efficiency of turbulent mixing. This long distance nonequilibrium transport is distinguishable from the traditional turbulent models, such as eddy viscosity-type RANS and LES, where only turbulent viscosity coefficients are highly enhanced and the equilibrium diffusion process are remained in almost all previous turbulence modeling.

For the turbulent modelling, at elevated Reynolds numbers, intense strain rate induces the fragmentation of fluid structure into discrete elements through viscous destabilization. This phenomenon manifests as a breakdown of spatial correlations in the flow field, leading to the formation of isolated fluid parcels that exhibit quasi-particle behavior. These detached fluid elements, henceforth referred to as "fluid particles"  undergo autonomous transport while maintaining their individual characteristics. These fluid particles exist and evolve within a continuous background field characterized by smooth wave-like distributions of velocity, pressure, and other fluid properties. This duality creates a hierarchical structure where discrete fluid particles traverse and interact with an underlying continuous wave field. The background field exhibits conventional hydrodynamic wave propagation characteristics and provides the mean flow framework within which the particles operate. The particles, in turn, can be viewed as perturbations or fluctuations superimposed on this continuous substrate. Figure \ref{Fig-wp} presents such a picture for the wave-particle decomposition and evolution in a discretized space with the laminar and turbulent distribution.

\begin{figure}[htbp]
	\centering
	\subfigure[]{
		\includegraphics[height=7.0cm]{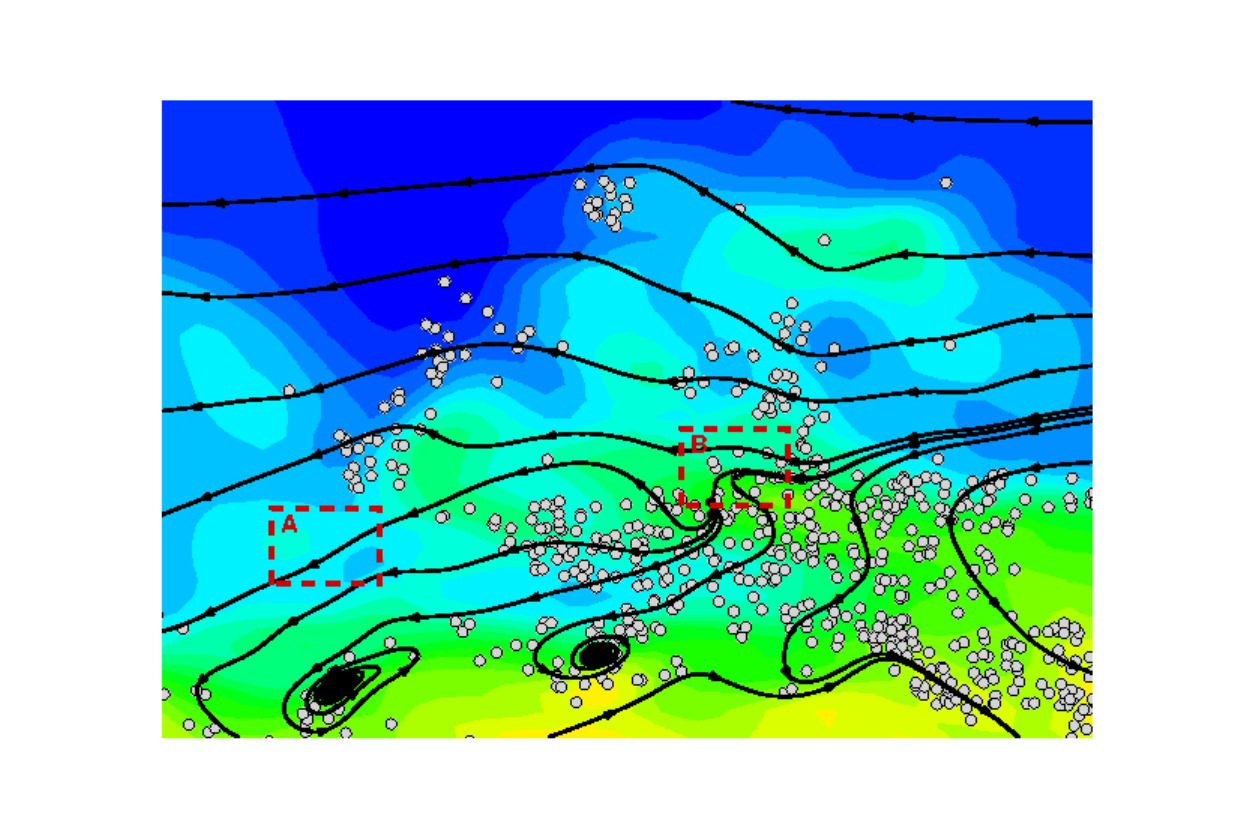}
	}
	\\
	\subfigure[]{
		\includegraphics[height=4.7cm]{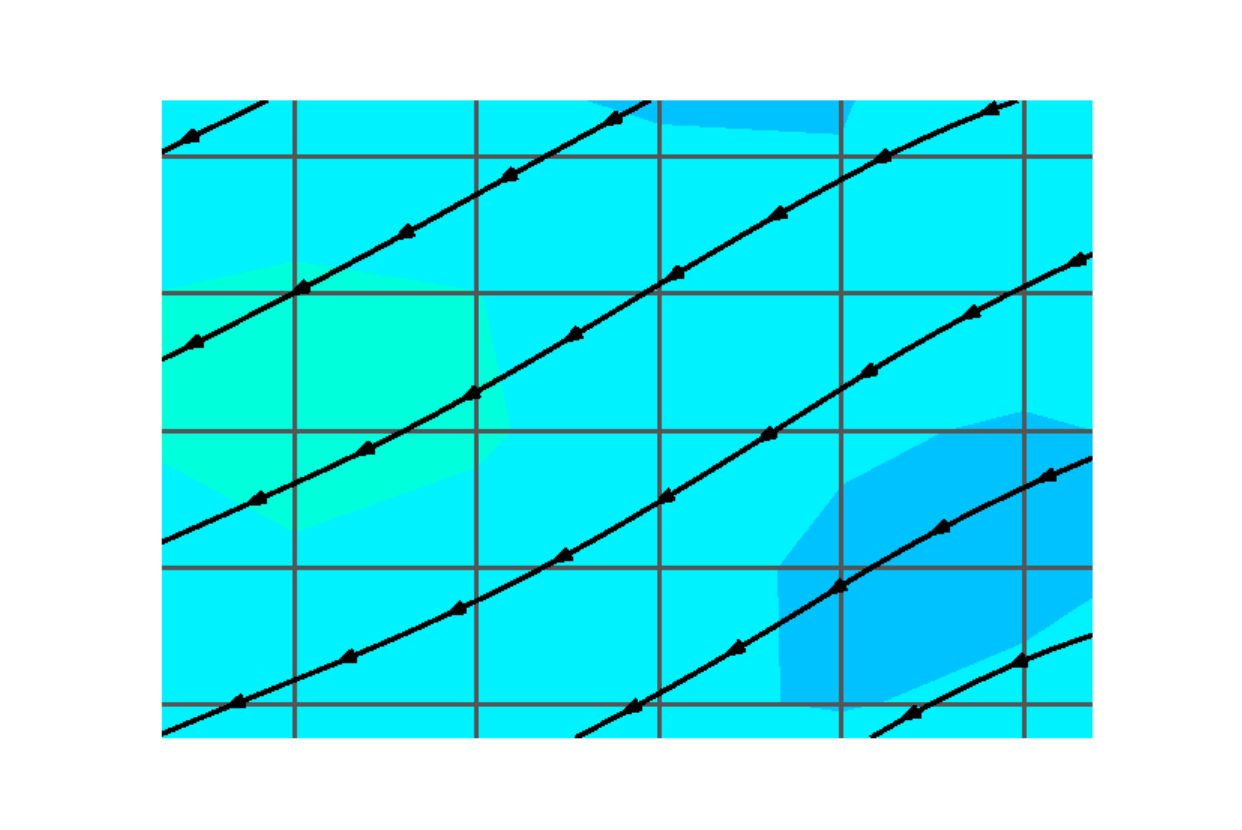}
	}
	\subfigure[]{
		\includegraphics[height=4.7cm]{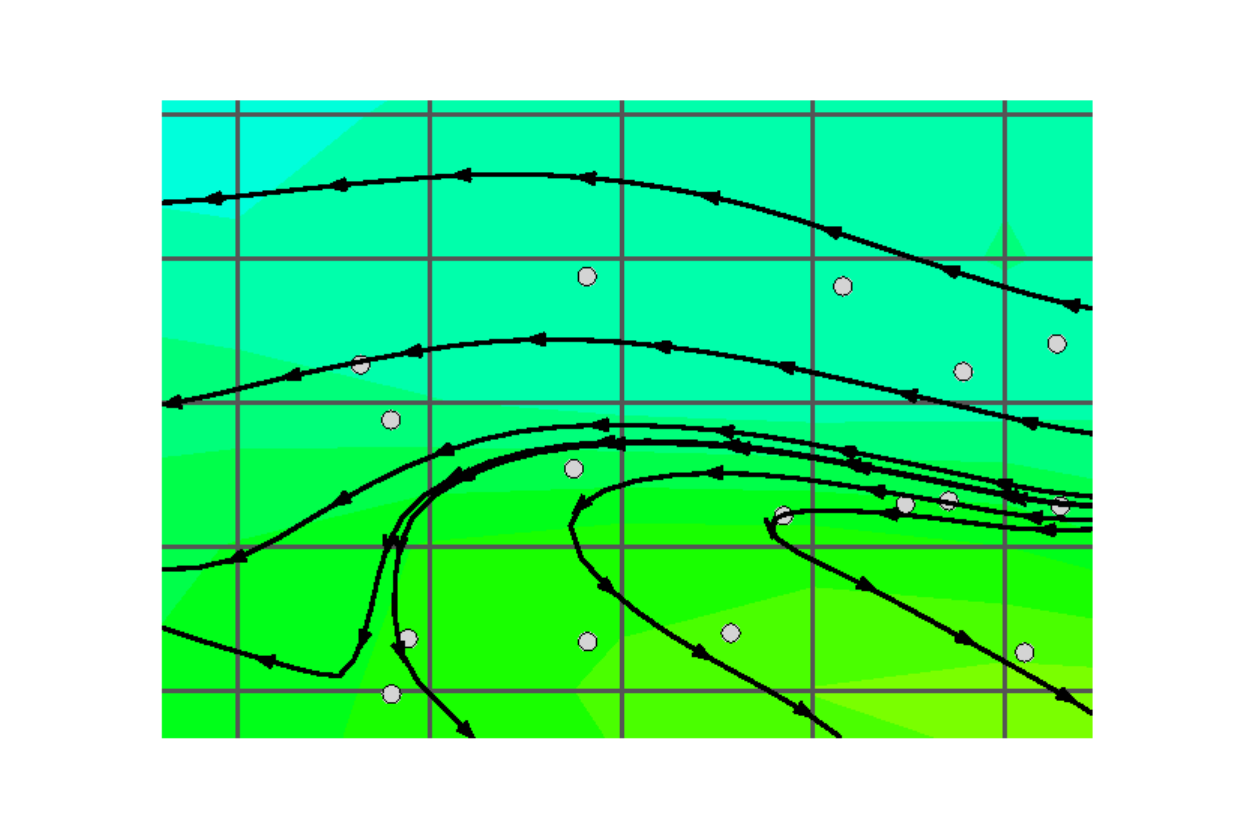}
	}
	\caption{The wave particle representation of turbulent flow in a discretized space with mesh size resolution. The background is contoured by streamwise velocity. (b) and (c) are the enlarged figure in zone A and zone B as shown in (a), respectively. The grid size is shown in (b) and (c) which is used in the WPTS.}
	\label{Fig-wp}
\end{figure}

The dynamic interaction between the particulate elements and the background wave field creates a rich multiscale phenomenon. While the background field evolves according to traditional continuum mechanics principles, the fluid particles respond to both the local gradients in the background field and their own inertial characteristics. This leads to a complex coupling where the particles can both influence and be influenced by the underlying wave field. The background field serves as a mediating mechanism for long-range interactions between particles, while the particles themselves contribute to local non-equilibrium transport phenomena and energy cascade processes.

This wave-particle duality provides a natural framework for understanding turbulent transport mechanisms. The continuous wave component captures large-scale coherent structures and pressure-driven phenomena, while the particle representation accounts for intense local mixing, non-equilibrium transport, and small-scale dissipative processes, etc. The relative dominance of wave-like or particle-like behavior varies spatially and temporally depending on local flow conditions, allowing for a smooth transition between varying flow regimes and turbulence intensities. Moreover, this decomposition offers insights into the mechanics of turbulent energy cascade. The interaction between fluid particles and the background wave field facilitates both forward and inverse energy cascades, with particles serving as primary agents for energy transfer across scales. The statistical properties of these particle-wave interactions emerge as fundamental determinants of bulk turbulent behavior, while preserving the essential non-local and non-equilibrium characteristics of turbulent flows.

In order to capture the above wave-particle decomposition of the turbulent flow, the UGKWP framework will be used to construct the dynamics model and simulate the turbulent flow. The fundamental concept of capturing equilibrium waves and non-equilibrium particles in the UGKWP method can be extended here with further inclusion of turbulent features. The significance for turbulence flow modeling is its wave-particle dual representation capabilities. The particles here represent discrete fluid elements transport under background wave-represented flow field, instead of real particle in the original UGKWP method for rarefied flow. This wave-particle decomposition for turbulent flow enables simultaneous evolution handling of large-scale coherent wave structures and small-scale turbulent particle eddies. Unlike traditional turbulence models relying solely on the diffusive processes with enhanced turbulent viscosity coefficients, this method captures genuine non-equilibrium transport through particle penetration in recovering large dissipative and mixing effects.

The method's adaptive capability automatically balances particle and wave representations based on local flow conditions, proving particularly valuable in simulating transitional flows. The UGKWP naturally captures laminar-turbulent transitions through replenishing and annihilating of particles,  controlled by the turbulence Knudsen number - the ratio between fluid element transport collision time and numerical time step. The UGKWP methodology transcends conventional turbulence model limitations by directly incorporating non-local, non-equilibrium multiscale transport mechanisms. The variable degrees of freedom in turbulent flow representation naturally unify laminar and turbulent flow description, pioneering new approaches to turbulent flow beyond the traditional Navier-Stokes flow representation.

The particle representation accounts for non-equilibrium transport and intense local mixing, while the wave component describes large-scale coherent motion and pressure-driven phenomena. Their coupled evolution, mediated by local flow conditions, offers a comprehensive description of turbulent flow physics across scales and regimes. The UGKWP framework will automatically recover the GKS for the Navier-Stokes solution in the laminar flow regime due to the absence of particles. The dynamic evolution of laminar, transition, and turbulent flow is constructed under a single wave-particle framework. In the next section, the UGKWP method for the turbulent flow will be presented.

\section{Wave-particle turbulent simulation method}	

\subsection{Kinetic model for turbulent flow}

The kinetic model equation for turbulent flow will adopt the BGK relaxation model
\begin{gather}\label{bgk}
\frac{\partial f}{\partial t}
+ \nabla_x \cdot \left(\vec{u}f\right)
= \frac{g-f}{\tau},
\end{gather}
where $f$ is the probability distribution function (PDF) of molecules, $\vec{u}$ is the velocity of molecules, $\tau$ is the collision time, and $g$ is the equilibrium state with
\begin{gather}\label{geq}
g = \rho \left(\frac{\lambda}{\pi} \right)^{\frac{K+3}{2}} e^{-\lambda \left[\left(\vec{u} - \vec{U}\right)^2 + \vec{\xi}^2 \right]}.
\end{gather}
$K$ is the internal degree of freedom taken as 2 for diatomic molecule gas. $\lambda = \frac{1}{2RT}$ and usually $T$ is the thermal temperature.
Further the conservative variables $\vec{W}=\left(\rho, \rho \vec{U}, \rho E\right)^T$ can be obtained by taking moment with $g$ by $\vec{\psi}$, where $\vec{\psi}=(1,\vec{u},\displaystyle \frac{1}{2}\left(\vec{u}^2+\vec{\xi}^2\right))^T$.
The conservation law is satisfied based on the compatibility condition
\begin{gather*}
\int \vec{\psi}\left(g - f\right)\text{d}\vec{\Xi} = \vec{0},
\end{gather*}
where $\text{d}\vec{\Xi}=\text{d}u\text{d}v\text{d}w\text{d}\xi_1...\text{d}\xi_{K}$.

The above kinetic relaxation model is a model about the relaxation process from the turbulent to laminar flow.
For turbulent flow, the molecule here is referred to the hydrodynamic particle, which is the discrete fluid element, as stated in the last section.
The gas distribution function $f$ is a fully non-equilibrium distribution for the turbulent flow. The solution from the kinetic model recovers a fully non-equilibrium transport process. In other words, the kinetic equation itself doesn't correspond to the quasi-equilibrium Navier-Stokes equations.
To model the non-equilibrium transport and use it in the numerical simulation are the essential ingredients for the construction of the new turbulent modeling and simulation in this paper, especially under the coarse mesh resolution.
In the equilibrium state, for the turbulence simulation the $\lambda$ in Eq.\eqref{geq} includes both the thermal temperature $T_{thermal}$ and the turbulence temperature $\Theta_t$, where the turbulent temperature is used to recover the unresolved turbulence kinetic energy (TKE), denoted by $\rho E_t$.
As a result, the total energy $\rho E$ in the current model includes not only the kinetic energy $\frac{1}{2}\rho \vec{U}^2$, thermal energy $\rho R T_{theraml}$, and  the TKE $\rho E_t$ as well. Correspondingly, the compatibility condition ensures the total energy $\rho E$ conservation, and the energy exchange among them determines the dynamics of turbulent flow, such as forward and inverse cascade. The energy distributions among kinetic, thermal, and turbulent are given in Appendix A.
Besides the energies, modeling the collision time $\tau$ also plays an important role for the turbulent flow. Since the kinetic model is used to simulate the turbulent flow without fully resolving the flow structure. Inside each cell, the multiple flow structure exist and transport. These unresolved flow structures are coming from the breaking down of the connetion of fluid elements. Under the cell size resolution $\Delta$ and the strain rate $\bold S$ in the current cell, it is assumed that the real physical strain rate ${\bold S}_{phys}$, resolved by the DNS resolution $\Delta_{DNS}$,  should be ${\bold S}_{phys} = {\bold S}_{DNS} = \sim (\Delta /\Delta_{DNS})^\alpha {\bold S}$. The collision time $\tau$ in the kinetic model will depend on the magnitude $\Delta_{DNS}$. The detailed formula for $\tau$ will be presented when the costruction of the numerical scheme for the turbulent flow. The non-equilibrium distribution function $f$ will be decomposed into the wave-particle components, where the wave propagates by the Navier-Stokes type laminar flow evolution and the particle takes non-equilibrium transport for the turbulent part. The emerging and disappearing of the particles is associated with the transition between the laminar and turbulent flow.

For the kinetic model equation, the integral solution of Eq.\eqref{bgk} is
\begin{equation}\label{bgk-integrasol}
f(\vec{x},t,\vec{u})=\frac{1}{\tau}\int_0^t g(\vec{x}',t',\vec{u} )e^{-(t-t')/\tau}\text{d}t'\\
+e^{-t/\tau}f_0(\vec{x}-\vec{u}t, \vec{u}),
\end{equation}
where $\vec{x}'=\vec{x}+\vec{u}(t'-t)$ is the trajectory of particles, $f_0$ is the initial gas distribution function at time $t=0$.

Generally, in the finite volume framework (FVM), the cell-averaged macroscopic variables $\vec{W}_i$ of cell $i$ can be updated by the conservation law,
\begin{gather}
\vec{W}_i^{n+1} = \vec{W}_i^n - \frac{\Delta t}{\Omega_i} \sum_{S_{ij}\in \partial \Omega_i}\vec{F}_{ij}S_{ij},
\end{gather}
where $\vec{W}_i=\left(\rho_i, \rho_i \vec{U}_i, \rho_i E_i\right)^T$ is the cell-averaged macroscopic variables,
\begin{gather*}
\vec{W}_i = \frac{1}{\Omega_{i}}\int_{\Omega_{i}} \vec{W}\left(\vec{x}\right) \text{d}\Omega,
\end{gather*}
$\Omega_i$ is the volume of cell $i$, $\partial\Omega_i$ denotes the set of cell interfaces of cell $i$, $S_{ij}$ is the area of the $j$-th interface of cell $i$, $\vec{F}_{ij}$ denotes the macroscopic fluxes (in unit time) across the interface $S_{ij}$, which can be written as
\begin{align}\label{particle phase Flux equation}
\vec{F}_{ij}= \frac{1}{\Delta t} \int_{0}^{\Delta t} \int \vec{u}\cdot\vec{n}_{ij} f_{ij}(\vec{x},t,\vec{u}) \vec{\psi} \text{d}\vec{u}\text{d}t,
\end{align}
where $\vec{n}_{ij}$ is the normal unit vector of interface $S_{ij}$, $f_{ij}\left(t\right)$ is the time-dependent distribution function on the interface $S_{ij}$.
Substituting the time-dependent distribution function Eq.\eqref{bgk-integrasol} into Eq.\eqref{particle phase Flux equation}, the fluxes can be obtained,
\begin{align}\label{free-transport}
\vec{F}_{ij}
&=\frac{1}{\Delta t} \int_{0}^{\Delta t} \int \vec{u}\cdot\vec{n}_{ij} f_{ij}(\vec{x},t,\vec{u}) \vec{\psi} \text{d}\vec{u}\text{d}t \nonumber \\
&=\frac{1}{\Delta t} \int_{0}^{\Delta t} \int\vec{u}\cdot\vec{n}_{ij} \left[ \frac{1}{\tau}\int_0^t g(\vec{x}',t',\vec{u})e^{-(t-t')/\tau}\text{d}t' \right] \vec{\psi} \text{d}\vec{u}\text{d}t \nonumber \\
&+\frac{1}{\Delta t} \int_{0}^{\Delta t} \int\vec{u}\cdot\vec{n}_{ij} \left[ e^{-t/\tau}f_0(\vec{x}-\vec{u}t,\vec{u})\right] \vec{\psi} \text{d}\vec{u}\text{d}t \nonumber \\
&\overset{def}{=}\vec{F}^{eq}_{ij} + \vec{F}^{fr}_{ij}.
\end{align}

Based on the above flux transport, the FVM scheme can be developed for the updates of conservative flow variables.
For the well-resolved laminar flow simulation, the corresponding gas-kinetic scheme (GKS) has been developed for the NS solution.
For the unresolved turbulent flow, the wave-particle turbulent simulation (WPTS) method will be constructed. In the following, the constructions of GKS and WPTS will be presented. The GKS is basically the limiting scheme of WPTS in the laminar flow region.

\subsection{GKS for the NS solution}

For the simulation under well-resolved grid, such as the laminar flow with zero TKE, the GKS as a NS solver will be presented \cite{GKS-2001} for the reference of WPTS. For GKS, the TKE $\rho E_t$ is zero. In other words, the GKS will become the limiting scheme of WPTS in the region without particles with zero TKE.

In GKS, for a second-order accuracy, the equilibrium state $g$ around the cell interface is written as,
\begin{gather*}
g\left(\vec{x}',t',\vec{u}\right)=g_0\left(\vec{x},\vec{u}\right)
\left(1 + \overline{\vec{a}} \cdot \vec{u}\left(t'-t\right) + \bar{A}t'\right),
\end{gather*}
where $\overline{\vec{a}}=\left[\overline{a_1}, \overline{a_2}, \overline{a_3}\right]^T$, $\overline{a_i}=\frac{\partial g}{\partial x_i}/g$, $i=1,2,3$,  $\overline{A}=\frac{\partial g}{\partial t}/g$, and $g_0$ is the local equilibrium on the interface.
Specifically, the coefficients of spatial derivatives $\overline{a_i}$ can be obtained from the corresponding derivatives of the macroscopic variables,
\begin{equation*}
\left\langle \overline{a_i}\right\rangle=\partial \vec{W}_0/\partial x_i,
\end{equation*}
where $i=1,2,3$, and $\left\langle...\right\rangle$ means the moments of the Maxwellian distribution functions,
\begin{align*}
\left\langle...\right\rangle=\int \vec{\psi}\left(...\right)g\text{d}\vec{u}.
\end{align*}
The coefficients of temporal derivative $\overline{A}$ can be determined by the compatibility condition,
\begin{equation*}
\left\langle \overline{\vec{a}} \cdot \vec{u}+\overline{A} \right\rangle = \vec{0}.
\end{equation*}
Now, all the coefficients in the equilibrium state $g\left(\vec{x}',t',\vec{u}\right)$ have been determined, and its integration becomes,
\begin{gather}\label{eqfeq}
f^{eq}(\vec{x},t,\vec{u}) \overset{def}{=} \frac{1}{\tau}\int_0^t g(\vec{x}',t',\vec{u})e^{-(t-t')/\tau}\text{d}t' \nonumber\\
= c_1 g_0\left(\vec{x},\vec{u}\right)
+ c_2 \overline{\vec{a}} \cdot \vec{u} g_0\left(\vec{x},\vec{u}\right)
+ c_3 A g_0\left(\vec{x},\vec{u}\right),
\end{gather}
with coefficients,
\begin{align*}
c_1 &= 1-e^{-t/\tau}, \\
c_2 &= \left(t+\tau\right)e^{-t/\tau}-\tau, \\
c_3 &= t-\tau+\tau e^{-t/\tau},
\end{align*}
and thereby the integrated flux over a time step for the equilibrium state can be obtained,
\begin{gather}\label{eqFluxeq}
\vec{F}^{eq}_{ij}
=\frac{1}{\Delta t} \int_{0}^{\Delta t} \int \vec{u}\cdot\vec{n}_{ij} f_{ij}^{eq}(\vec{x},t,\vec{u})\vec{\psi}\text{d}\vec{u}\text{d}t.
\end{gather}

The initial gas distribution function $f_0^k$, $k=l,r$, is constructed as
\begin{equation*}
f_0^k=g^k\left(1+a^kx+b^ky+c^kz-\tau(a^ku+b^kv+c^kw+A^k)\right),
\end{equation*}
where $g^l$ and $g^r$ are the Maxwellian distribution functions on the left and right-hand sides of a cell interface, and they can be determined by the corresponding conservative variables $\vec{W}^l$ and $\vec{W}^r$. The coefficients $a^l$, $a^r$, $b^l$, $b^r$, $c^l$, $c^r$ are related to the spatial derivatives in normal and tangential directions, which can be obtained from the corresponding derivatives of the initial macroscopic variables,
\begin{gather*}
\left\langle a^l\right\rangle=\partial \vec{W}^l/\partial x,
\left\langle a^r\right\rangle=\partial \vec{W}^r/\partial x,\\
\left\langle b^l\right\rangle=\partial \vec{W}^l/\partial y,
\left\langle b^r\right\rangle=\partial \vec{W}^r/\partial y,\\
\left\langle c^l\right\rangle=\partial \vec{W}^l/\partial z,
\left\langle c^r\right\rangle=\partial \vec{W}^r/\partial z.
\end{gather*}

The non-equilibrium parts of the Chapman-Enskog expansion have no net contribution to the conservative variables,
\begin{equation*}
\left\langle a^lu+b^lv+c^lw+A^l\right\rangle = 0,~
\left\langle a^ru+b^rv+c^rw+A^r\right\rangle = 0,
\end{equation*}
and therefore the coefficients $A^l$ and $A^r$, related to time derivatives, can be obtained. Then we have,
\begin{gather}\label{eqffr}
f^{fr}(\vec{x},t,\vec{u}) \overset{def}{=} e^{-t/\tau}f_0(\vec{x}-\vec{u}t,\vec{u}) \nonumber\\
= c_4 g^k\left(\vec{x},\vec{u}\right)
+ c_5 \vec{a}^k \cdot \vec{u} g^k\left(\vec{x},\vec{u}\right)
+ c_6 A^k g^k\left(\vec{x},\vec{u}\right),
\end{gather}
with coefficients,
\begin{align*}
c_4 &= e^{-t/\tau}, \\
c_5 &= -\left(t+\tau\right)e^{-t/\tau}, \\
c_6 &= -\tau e^{-t/\tau},
\end{align*}
and thereby the integrated flux over a time step for the equilibrium state can be obtained,
\begin{gather}\label{eqFluxfr}
\vec{F}^{fr}_{ij}
=\frac{1}{\Delta t} \int_{0}^{\Delta t} \int \vec{u}\cdot\vec{n}_{ij} f_{ij}^{fr}(\vec{x},t,\vec{u})\vec{\psi}\text{d}\vec{u}\text{d}t.
\end{gather}

Finally, the time-dependent distribution function $f(\vec{x},t,\vec{u})$ at a cell interface can be expressed as,
\begin{align}\label{eqffullgks}
f(\vec{x},t,\vec{u})=&(1-e^{-t/\tau})g_0\left(\vec{x},\vec{u}\right)+[(t+\tau)e^{-t/\tau}-\tau]\overline{\vec{a}} \cdot \vec{u} g_0\left(\vec{x},\vec{u}\right)\nonumber\\
+&(t-\tau+\tau e^{-t/\tau}){\bar{A}} g_0\left(\vec{x},\vec{u}\right)\nonumber\\
+&e^{-t/\tau}[1-(\tau+t)\vec{a}^r \cdot \vec{u}-\tau A^r]g^r\left(\vec{x},\vec{u}\right)(1-H(u))\nonumber\\
+&e^{-t/\tau}[1-(\tau+t)\vec{a}^l \cdot \vec{u}-\tau A^l]g^l\left(\vec{x},\vec{u}\right)H(u).
\end{align}
This is the so-called GKS method \cite{GKS-2001}.
In this paper, the high-order GKS is employed for the turbulence simulation, and more details will be introduced later.


In GKS, the collision time $\tau$ is defined by $\tau={\mu}/{p}$ for the laminar flow computation, by which the NS solution with $Pr=1.0$ can be recovered.
The $\mu$ stands for the physical viscosity, which is obtained by the widely-used models, such as the power-law model.
In order to capture shock as the conventional shock capturing schemes, a numerical collision time $\tau_n$ is added to increase dissipation through the enhancement of the contribution from $\vec{F}^{fr}$ with the reduction of the percentage of $\vec{F}^{eq}$ \cite{GKS-2001,ji2024twostep,zhao2023high}.

\subsection{Wave-particle turbulent simulation method}

Once the turbulent flow cannot be well-resolved by the grid size, the flow structure and evolution inside each cell have to be modeled.
The un-resolved flow structure in the coarse mesh is associated with an increasing of degrees of freedom, such as the velocity difference between the local fluid element and the cell-averaged values. The traditional turbulent modeling is to mimic the dynamics of subcell structure and evolution by the cell averaged variables.
As a result, only the quasi-equilibrium models can be developed, such as the design of the turbulent viscosity coefficient and the dynamics is still controlled by the NS-type equilibrium process.
In the current WPTS model, the subcell flow structure within the coarse mesh is modeled and evolved as a combination of continuous wave propagation of cell averaged macroscopic variables through the NS dynamics and the discrete fluid elements movement through the non-equilibrium particle transport.
The non-equilibrium stochastic particle (fluid element) transport plays a dominant role  in recovering the dynamic features of the turbulent flow.

In WPTS, the distribution function $f$ is decomposed into wave and particle.
The coupled evolution of wave and particle will follow the numerical algorithm of unified gas-kinetic wave-particle (UGKWP) method.
The specialities for turbulent flow from the original UGKWP are the definition of turbulent collision time $\tau_t$, the distribution of wave and particle from macroscopic flow variables, and the particle transport and its annihilation and replenishing in the transport process in recovering the exchange between the turbulent kinetic energy (TKE) and the background wave energy. In the following, we will present the formulations in WPTS for the updates of macroscopic flow variables and the turbulent kinetic energy.

In the construction WPTS algorithm, we are considering the general case where the cell resolution is not enough to resolve the DNS flow structure.
In other words, the cell size is much larger than the DNS cell size. Under such a situation, like the LES model,
the turbulence collision time $\tau_t$ is usually much larger than the physical one determined by the viscosity coefficient of the NS equations $\tau = \mu /p$.
In the strong turbulent region, more particles with long distance transport for the efficient mixing will be modeled according to the particle collision time
$\tau_n = \tau + \tau_t$.  With the increasing flux from particle free transport ${\vec F}_{ij}^{fr}$ in Eq.\eqref{free-transport}, the contribution from the equilibrium flux ${\vec F}_{ij}^{eq}$ in Eq.\eqref{eqFluxeq}, which targets recovering the laminar NS solution, will be decreasing systematically. The details of the calculation of $\tau_t$ will be introduced later. Note the main difference between GKS and WPTS is the construction of the free transport part $f_0 ({\vec x}-{\vec u}t, {\vec u})$ of the integral solution of the kinetic relaxation model.
The local TKE can be directly obtained by counting the existing stochastic particles. The details of evaluating TKE can be found in Appendix B.

Different from the UGKWP for the rarefied flow, the WPTS method considers the update of turbulent kinetic energy and uses the specially designed relaxation time.
But the algorithms of WPTS and UGKWP have close similarities.
At the beginning of each time step, in WPTS the amount of conservative variables $\vec{W}$ are composed of
the particle part $\vec{W}^p$ for the turbulent kinetic energy and the wave part $\vec{W}^h$ for the background flow field.
In each time step, as shown in this section the total conservative flow variables $\vec{W}$ and the particle $\vec{W}^p$ will be updated inside each cell.
There, the wave part $\vec{W}^h$ can be obtained directly based on the conservation in each cell, namely,
\begin{gather}
	\vec{W}^{h}_i = \vec{W}_i - \vec{W}^p_i.
\end{gather}
During the evolution process, the amounts of mass, momentum, and energy from the particle and wave parts can be transferred in both directions to model the transition between the turbulent and laminar flow.
In the following, we are going to present the WPTS algorithm.

The WPTS is based on the evolution solution of the kinetic relaxation model Eq.\eqref{bgk-integrasol}.
The free transport part $f_0$ in the equation is mainly evaluated from the particle transport, which will be presented in the following.
The stochastic particle from $\vec{W}^p$ mimics the discrete fluid element and its movement in the background fluid is modeled as
\begin{gather}\label{dudtpar}
\frac{\text{d} \vec{u}\left(t\right)}{\text{d} t} = \frac{\vec{U} - \vec{u}}{\tau_n} + \vec{a}.
\end{gather}
The particle movement is associated with the non-equilibrium of $f$.
The cell-averaged TKE comes from the particle stochastic movement with its individual velocity. As a result, the first term on the RHS of Eq.\eqref{dudtpar} indicates the relaxation to the local cell-resolved velocity $\vec{U}$ from the particle's discrete velocity $\vec{u}$. The second term on the RHS of Eq.\eqref{dudtpar} models the forcing term on the fluid element moving in a background flow field. Since the particle here is the broken hydrodynamic fluid element, its movement is more or less
controlled by the viscous stress in the macroscopic level. As a simple model, here only the pressure gradient force is considered, namely, $\vec{a} = \frac{\nabla p}{\rho}$.

For the relaxation term, an analytical solution can be obtained
\begin{gather*}
\vec{u}\left(t\right) = e^{-t/\tau_n} \vec{u}\left(t=0\right) + \left(1 - e^{-t/\tau_n}\right) \vec{U},
\end{gather*}
which means the stochastic fluid element has a probability of $e^{-t/\tau_n}$ to transport freely while $(1-e^{-t/\tau_n})$ to collide with other fluid elements and background fluid, thus merge into the local macroscopic flow in the process to evolve to the equilibrium state, or the laminar flow. The decaying process due to particle collision indicates the dissipation of TKE in the evolution. The free transport time before colliding with other fluid elements is denoted as $t_c$, and the cumulative distribution function of $t_c$ is,
\begin{gather}\label{particle phase wp cumulative distribution}
F\left(t_c < t\right) = 1 - e^{-t/ \tau_n}.
\end{gather}
Therefore, $t_c$ can be sampled as $t_c=-\tau_n \text{ln}\left(\eta\right)$, where $\eta$ is a random number generated from a uniform distribution $\mathcal{U}\left(0,1\right)$. Then, the free streaming time $t_f$ for each particle is determined individually by,
\begin{gather}
t_f = \text{min}\left[-\tau_n\text{ln}\left(\eta\right), \Delta t\right],
\end{gather}
where $\Delta t$ is the time step.
More specifically, the particle trajectory in the free streaming process within time $t_f$ is tacked by
\begin{gather}
\vec{x}^* = \vec{x}^n + \vec{u}^n t_f .
\end{gather}
With the inclusion of particle acceleration in the transport process, both velocity $\vec{u}$ and the position $\vec{x}$ of the particle ia updated as,
\begin{align}
\vec{u}^{n+1} &= \vec{u}^n + \vec{a} t_f, \\
\vec{x}^{n+1} &= \vec{x}^* + \frac{1}{2}\vec{a} t_f^2.
\end{align}
Up to now the tracking of the particles for the evolution of their accumulated $\vec{W}^p$ is finished.
In summary, the trajectories of these particles have been tracked in a time interval $\left(0, t_f\right)$.

For the transport particles with $t_f=\Delta t$, they will survive at the end of a one-time step. However, the collisional particles with $t_f<\Delta t$ are deleted after tracking up to $t_f$.
In the above process, the non-equilibrium transport of fluid element is basically captured by the stochastic particle.
On the other hand, a fraction of stochastic particles will be eliminated and merged into the background flow inside the updated total $\vec{W}$.
This particle eliminating process is associated turbulent dissipation.
In other words, $t_f$ for the stochastic particle describes a relaxation of turbulence flow to a locally resolved large-scale laminar flow structure, indicating the dynamic dissipation of TKE in the particle collisional process.
In turbulence study, it is well recognized that, besides the dissipation, the production of TKE is also important.
In WPTS, the production of TKE is complemented from the newly sampled particles from $\vec{W}^h$, which is determined from the updated $\vec{W}$ and $\vec{W}^p$, as shown later.
In particular, the velocity fluctuation of newly sampled particles is assumed to have a normal distribution and more specifically determined by $\delta\vec{u}_p = \mathcal{D}_{N} \left[C_0\left(1-e^{\Delta t/\tau_n}\right)\rho E_t, \rho^h\right]$, where $C_0$ is the coefficient. Please refer to Appendix C for the definition of $\mathcal{D}_{N}$ and the details of the sampling process.


As described above, a particular feature of the stochastic particle is that it has a probability of $e^{-t/\tau_n}$ to free transport and $(1-e^{-t/\tau_n})$ to merge into the local background flow. In order to simulate the flow efficiently, same as the approach in UGKWP, since the collisional particles in the next time step will be eliminated and at the beginning of next time step only collisionless paricles from $\vec{W}^h$ need to sample from the amount of macroscopic variables
\begin{gather}
\vec{W}^{hp}_i = e^{-\Delta t/\tau_n} \vec{W}^{h}_i,
\end{gather}
and the free transport time of these newly sampled stochastic particles from $\vec{W}^{hp}$ is taken as $t_f=\Delta t$. These collisionless particles will be tracked as well in the next time step without collision. But, the remaining particles from the previous time step will transport with the possible collision described above.
The transport of un-sampled particles from $(\vec{W}^{h} - \vec{W}^{hp})$ can be calculated analytically.
Their contribution in the free transport $f_0$ to the flux, denoted as $\vec{F}^{fr,wave}$, is evaluated analytically \cite{WP-first-liu2020unified, WP-second-zhu-unstructured-mesh-zhu2019unified, UGKS-book-framework-xu2021cambridge}
\begin{align*}\label{eqFluxfrwave}
\vec{F}^{fr,wave}_{ij}
&=\frac{1}{\Delta t}\int_{0}^{\Delta t} \int \vec{u} \cdot \vec{n}_{ij} \left[ e^{-t/\tau_n}f_0(\vec{x}-\vec{u}t,\vec{u})\right] \vec{\psi} \text{d}\vec{u}\text{d}t\\
&-e^{-\Delta t/\tau_n} \frac{1}{\Delta t} \int_{0}^{\Delta t} \int \vec{u} \cdot \vec{n}_{ij} \left[g_0^h\left(\vec{x},\vec{u} \right) - t\vec{u} \cdot g_\vec{x}^h\left(\vec{x},\vec{u} \right) \right] \vec{\psi}\text{d}\vec{u}\text{d}t\\
&=\frac{1}{\Delta t} \int \vec{u} \cdot \vec{n}_{ij} \left[ \left(q_4  - \Delta t e^{-\Delta t/\tau_n}\right) g_0^h \left(\vec{x},\vec{u} \right)
+ \left(q_5 + \frac{\Delta t^2}{2}e^{-\Delta t/\tau_n}\right) \vec{u} \cdot g_\vec{x}^h\left(\vec{x},\vec{u} \right) \right]\vec{\psi}\text{d}\vec{u},
\end{align*}
with
\begin{gather*}
q_4=\tau_n\left(1-e^{-\Delta t/\tau_n}\right), \\
q_5=\tau_n\Delta te^{-\Delta t/\tau_n} - \tau_n^2\left(1-e^{-\Delta t/\tau_n}\right).
\end{gather*}
The transport flux at the cell interface from the stochastic particle can be obtained directly by counting particles.
For example, the flux of cell $i$ due to the transport of stochastic particles, denoted as $\vec{w}_{i}^{fr,part}$, is obtained by counting the particles passing through the interfaces of cell $i$,
\begin{gather}
\vec{w}_{i}^{fr,part} = \sum_{k\in P\left(\partial \Omega_{i}^{+}\right)} \vec{\phi}_k - \sum_{k\in P\left(\partial \Omega_{i}^{-}\right)} \vec{\phi}_k,
\end{gather}
where $P\left(\partial \Omega_{i}^{+}\right)$ is the particle set moving into the cell $i$ during one time step, $P\left(\partial \Omega_{i}^{-}\right)$ is the particle set moving out of the cell $i$ during one time step, $k$ is the particle index in the set, and $\vec{\phi}_k=\left[m_{k}, m_{k}\vec{u}_k, \frac{1}{2}m_{k}\vec{u}^2_k + m_k\frac{K+3}{2} \frac{1}{2\lambda_k}\right]^T$ is the mass, momentum and energy carried by particle $k$.
Note that, the conservative variables $\vec{\phi}_k$ carried by the sampled particle, $K=0$ is for the monatomic molecule and $K=2$ for the diatomic molecule. In this paper, the $K=2$ is employed, and correspondingly $\gamma = 1.4$.

Therefore, the updating of the cell-averaged total macroscopic variables can be written as
\begin{gather}\label{particle phase equ_updateW_ugkp}
\vec{W}_i^{n+1} = \vec{W}_i^n
- \frac{\Delta t}{\Omega_i} \sum_{S_{ij}\in \partial \Omega_i}\vec{F}^{eq}_{ij}S_{ij}
- \frac{\Delta t}{\Omega_i} \sum_{S_{ij}\in \partial \Omega_i}\vec{F}^{fr,wave}_{ij}S_{ij}
+ \frac{\vec{w}_{i}^{fr,part}}{\Omega_{i}}.
\end{gather}
The update of $\vec{W}_i^{p}$ inside each cell can be obtained by summing the contributions from all particles survived inside the cell.

The procedure of WPTS including the evolution of both wave and particle components is summarized in Figure \ref{Fig-wpts}.
\begin{figure}[htbp]
	\centering
	\includegraphics[height=8.0cm]{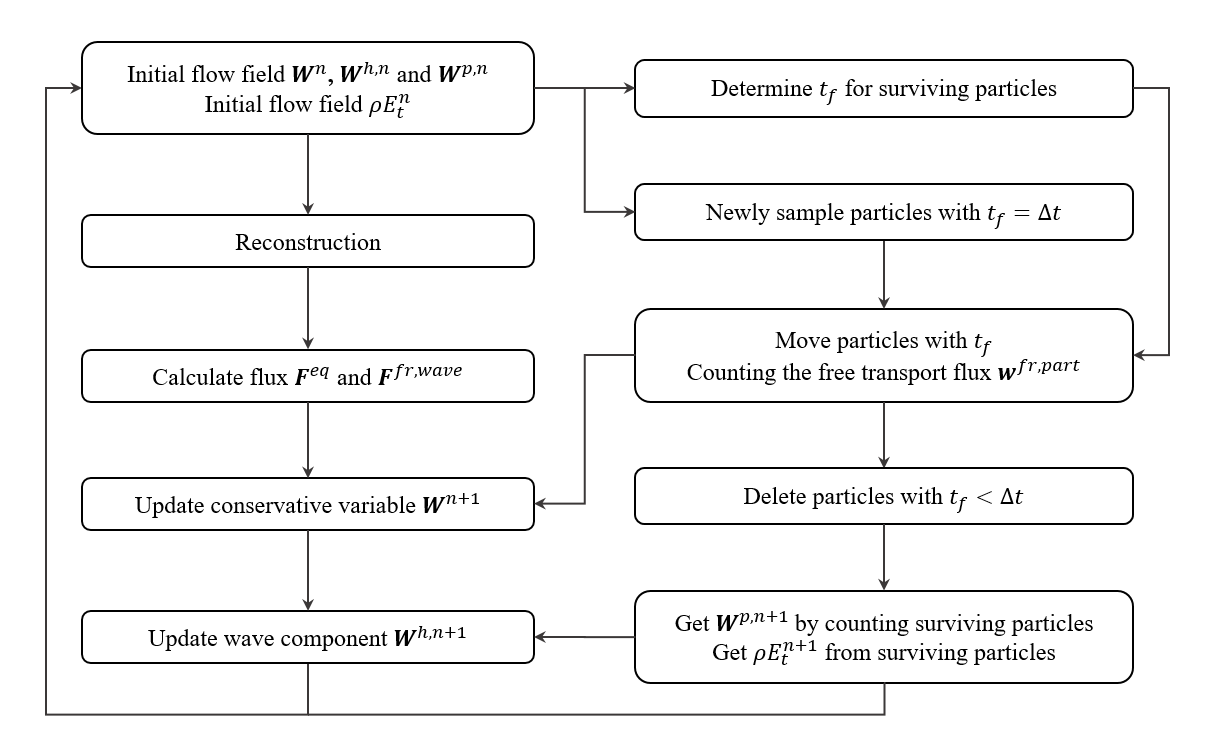}
	\caption{The procedure of WPTS.}
	\label{Fig-wpts}
\end{figure}

For the well-resolved flow structure by the cell size, with the absence of particles, the above WPTS will recover the GKS presented in the last sub-section exactly, which is a standard DNS solver.


The above algorithm WPTS is almost identical to UGKWP, but with the inclusion of TKE and modified particle trajectory.
However, for the rarefied flow a clear definition of particle collision time $\tau$ is known, such as the physical $\tau$ used in the GKS.
For the turbulent flow study, the critical parameter to determine the validity and quality of the modeling and simulation of turbulent flow is the
determination of turbulence collision time $\tau_t$, which is used in the above WPTS.
In this paper, as a first attempt, we assume that $\tau_t$ depends on both resolved macroscopic flow variables and existing particles,
\begin{align}\label{tautur}
e^{-\Delta t / \tau_t} = \left(1-\omega_p\right) e^{-\Delta t / \tau_{mac}} + \omega_p E_p,
\end{align}
where $\omega_p$ and $\left(1-\omega_p\right)$ are the weights for the stochastic particles and macroscopic flow structures, respectively,
such as $\omega_p = k \alpha_p$ and $\alpha_p = \rho^p / \rho$, where $\rho^p$ is the corresponding density from stochastic particles.
Here $k$ is the value around $1.0$, but limited to zero at small local velocity fluctuation. In this paper, it is taken as
\begin{gather}
k = \left\{
\begin{array}{ll}
0, & \sqrt{\Theta_t} < 10^{-3} ,\\
0.98, & \sqrt{\Theta_t} \ge 10^{-3}.
\end{array}
\right.
\end{gather}
$E_p$ indicates the non-equilibrium influence from the stochastic particles, namely the status of the local discrete fluid elements.
As the counterpart of the term $e^{-\Delta t / \tau_{mac}}$, here we assume $E_p \in \left[0,1\right]$, and obviously the larger value of $E_p$ will lead to the higher dependence of $\tau_t$ on the particle percentage $\alpha_p$. In this paper, the $E_p$ is taken as a constant value with 0.5 without further specification.

Besides, the determination of $\tau_{mac}$ is about the modeling of the number of discrete fluid element inside each cell. It is related to the cell size and the strength of the strain rate to break the connection between fluid elements. Here we adopt the widely used Smagorinsky model (SM), by which the local strain rate tensor $\vec{S}$ is used to determine the turbulent viscosity coefficient,
\begin{gather}\label{taumac}
\tau_{mac} = \frac{\mu_{SM}}{p} = \frac{C_s^2 \rho \Delta^2 |\vec{S}|}{p},
\end{gather}
where $\Delta$ is the cell size obtained by $\left(\Delta_x \Delta_y \Delta_z \right)^{1/3}$, and $C_s$ is the coefficient and in this paper we take $C_s^2 = 0.015$.

In LES, the filtering technique is used to reconstruct fluid dynamic equations under the cell size $\Delta$.
Even though the SM is employed  for the determination of $\tau_{mac}$ in WPTS, we have a different understanding for the SM model.
In our model, we need to know how many discrete fluid elements will be generated locally, and the rate of generation needs to be closely related to the real physical
strain rate tensor. However, numerically on the cell resolution $\Delta$, we only know the coarse-graining strain rate tensor $\vec S$. The real physical strain rate ${\vec S}^*$ can be only obtained in the DNS simulation. We believe that there should have a connection between $\vec S$ in mesh size $\Delta$ and ${\vec S}^*$ in DNS mesh size $\Delta_{DNS}$. With the assumption
$|\vec{S}^*| = \left(\frac{\Delta}{\Delta_{DNS}}\right)^2 |\vec{S}|,$
the definition of $\tau_{mac}$ can be reformulated as,
	\begin{gather}\label{taumacnew}
	\tau_{mac} = \frac{\mu_{mac}}{p} = \frac{C_s^2 \rho \Delta_{DNS}^2 |\vec{S}^*| }{p},
	\end{gather}
which is a physical parameter independent of mesh $\Delta$. In other words, intrinsically the WPTS is targeting on the dynamics of fully resolved turbulent flow, but from a much coarse mesh size $\Delta$. The increased degree of freedom through the particles is somehow to recover the DNS dynamics.
Overall, the modeling of turbulent collision time $\tau_{t}$ in Eq.\eqref{tautur} is a general model for capturing local turbulent non-equilibrium, incorporating information from both resolved large-scale flow structure and unresolved sub-grid flow dynamics.
More accurate turbulent modeling can be developed through the construction of $\tau_t$ with refined physical consideration.

\section{Numerical test}
In this section, the compressible mixing layer (CML) will be studied by using WPTS under an unresolved grid. The results from WPTS will be compared with the experimental measurement and DNS reference solutions. In order to validate WPTS, under the same coarse mesh, the NS solution from the fifth-order GKS and the LES solution from GKS will be obtained as well for comparison.

\subsection{Code validation}
In this paper, the high-order GKS is employed to get the NS solution. More specifically, the fifth-order WENO-AO reconstruction is used to achieve a high-order spatial accuracy, and a two-stage fourth-order strategy is adopted for the time marching \cite{GKS-HLLC-compare-yang2022comparison, wenoao-gks-ji2019-performance-enhancement}. The performance of this scheme, such as its accuracy and robustness, has been tested in our previous studies. Besides, the benchmark cases of turbulence, the compressible isotropic turbulence, and the three-dimensional Taylor-Green vortex problem have also been calculated and compared with the reference solution, validating the reliability of the high-order GKS method. The details of high-order GKS used in this paper can be found in \cite{GKS-HLLC-compare-yang2022comparison}.

In WPTS, the fifth-order WENO-AO reconstruction is adopted for $\vec{W}$, by which the flux of equilibrium state $\vec{F}^{eq}$ is determined; while the second-order reconstruction with van Leer limiter is employed for $\vec{W}^h$ by which the flux $\vec{F}^{fr,wave}$ can be obtained. For the temporal discretization, the wave component is evolved by the two-step fourth-order method, while the stochastic particle component is transported as described above in a whole time step $\Delta t$.
Besides it is worth noting that, for the cells without any stochastic particle in WPTS, the flux $\vec{F}$ from Eq.\eqref{eqffullgks} based on $\vec{W}$ will be used, indicating the WPTS will automatically go back to the exactly high-order GKS in these cells, such that the laminar flow solution is obtained.

\subsection{The compressible mixing layer}

The compressible mixing layer (CML) is a cornerstone problem in the turbulence research community \cite{Tur-case-mixing-review-yoder2015modeling}.
The flow features of CML are significantly influenced by the Reynolds number, Mach number, etc. Many researchers have extensively explored it by both numerical \cite{Tur-case-mixing-DNS-pantano2002study, Tur-case-mixing-DNS-biancofiore2014crossover} and experimental techniques
\cite{Tur-case-mixing-EXP-elliott1990compressibility, Tur-case-mixing-EXP-gruber1993three}. For example, the key features of the fully developed state, in this case, have been widely studied, including the momentum thickness, the turbulence kinetic energy, the profiles of the Reynolds stresses \cite{Tur-case-mixing-DNS-pantano2002study, Tur-case-mixing-DNS-zhang2019direct, Tur-model-NTRKM-cao2021non}.
This section utilizes the temporal evolved CML as the benchmark case for the validation of the proposed WPTS method and lays the groundwork for its application to more realistic and challenging turbulent problems.

\begin{figure}[htbp]
	\centering
	\includegraphics[height=5.5cm]{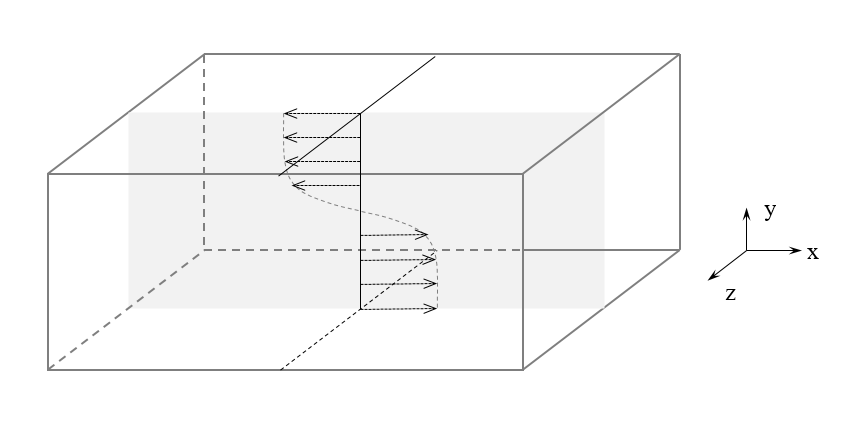}
	\caption{The sketch of domain and initial profile of mean streaming velocity.}
	\label{Fig-sketch}
\end{figure}

Figure \ref{Fig-sketch} shows the sketch of computational domain and $X\times Y\times Z$ is taken as $[0, 314.0]\times[-78.5, 78.5]\times[-39.25, 39.25]$.
The periodic boundary condition is used for left, right, back, and front boundaries; while the non-reflecting Riemann boundary condition is employed for the bottom and up boundaries.
The initial mean velocity field is given by
\begin{equation}
U_{\infty}\left(y\right) = \frac{1}{2} \Delta U \text{tanh}\left(-\frac{y}{2\delta_{\theta0}}\right),
~~~ V_{\infty}=0, ~~~ W_{\infty}=0,
\end{equation}
where $\delta_{\theta0}$ is the initial momentum thickness, taken as 1.0, and $\Delta U = U_l - U_u$ is the difference between the velocity of lower stream $U_l = 1.0$ and that of upper stream $U_u = -1.0$. The initial mean density is $\rho_{\infty} = 1.0$, and the initial mean temperature $T_{\infty}$ can be determined from
\begin{gather*}
Ma_c \overset{def}{=} \frac{\Delta U}{c_l + c_u} = \frac{\Delta U}{2 \sqrt{\gamma R T_{\infty}}},
\end{gather*}
with the given $Ma_c$ value.

To speed up turbulence onset, perturbations are superimposed onto the initial flow field, in addition to the mean field described above \cite{Tur-model-NTRKM-cao2021non}.
Particularly, the perturbation of flow variable $v$ is given by
\begin{gather}
v = v_{\infty} + A_{s1} \eta e^{-y^2 / \left(2 \delta_{\theta 0}\right)^2},
\end{gather}
where $v$ stands for $\rho$, $V$, $W$, and $T$. $\eta$ is the random number from a uniform distribution $\mathcal{U}\left[-0.5,0.5\right]$, and $A_{s1}$ is taken as 0.1 in this paper.
Then for the velocity in streamwise direction $U$, besides the above random perturbation, the following sinusoidal-type one is also added,
\begin{gather}
U = U_{\infty} + \left[A_{s1} + A_{s2} sin\left(\gamma_{s1} y\right) \left(B_{s1} + B_{s2}\right)\right] \eta e^{-y^2 / \left(2 \delta_{\theta 0}\right)^2},
\end{gather}
where
\begin{gather*}
B_{s1} = A_{s3} \left[cos\left(\gamma_{s2} x\right) + cos\left(2\gamma_{s2} x\right) + cos\left(4\gamma_{s2} x\right)\right], \\
B_{s2} = A_{s4} \left[cos\left(\gamma_{s2} x\right)cos\left(\gamma_{s2} z\right) + cos\left(2\gamma_{s2} x\right)cos\left(2\gamma_{s2} z\right) + cos\left(4\gamma_{s2} x\right)cos\left(4\gamma_{s2} z\right) \right],
\end{gather*}
and $A_{s2} = 0.6$, $A_{s3} = 0.2$, $A_{s4} = 0.4$, $\gamma_{s1} = 0.25$, $\gamma_{s2} = 0.235$.
The reference physical viscosity $\mu_{\infty}$ can be obtained from
\begin{gather*}
Re_{\omega 0} = \frac{\rho_{\infty} \Delta U \delta_{\omega 0}}{\mu_{\infty}},
\end{gather*}
with the given $Re_{\omega 0}$ value and $\delta_{\omega 0} = 4\delta_{\theta 0}$.
Then the physical viscosity is calculated by the power-law, namely, $\mu = \mu_{\infty} ({T}/{T_{\infty}})^{\omega}$ with $\omega = 0.667$.

One typical variable for the temporal turbulent mixing layer is the momentum thickness $\delta_{\theta}$, which is time-dependent with the development of turbulence flow structures and shows a self-similarity feature for the fully developed state. Specifically, it can be obtained by
\begin{gather*}
\delta_{\theta} = \frac{1}{\rho_{\infty} \left(\Delta U\right)^2} \int_{-\infty}^{\infty} \left(\overline{\rho U} - \overline{\rho} U_l\right) \left(\overline{\rho} U_u - \overline{\rho U}\right)\text{d}y,
\end{gather*}
where $\overline{v}$ means the spatial average in $xoz$ plane for variable $v$.
Furthermore, as the typical statistic variables for turbulence, the Reynolds stress terms $R_{ij}$ can be obtained by
\begin{gather}
R_{ij} = \overline{\rho \left(U_i - \widetilde{U}_i\right) \left(U_j - \widetilde{U}_j\right)} / \overline{\rho},
\end{gather}
where $\widetilde{v} = \overline{\rho v} / \overline{v}$ is the Favre average for variable $v$.

In this paper, the case with $Ma_c = 0.70$ and $Re_{\omega 0} = 640$ in \cite{Tur-case-mixing-DNS-pantano2002study} is simulated by WPTS.
This study employs uniform cells with number $144\times96\times48$, resulting in cell sizes 3.25, 2.45, and 2.45 times larger than those used in the DNS simulation in \cite{Tur-case-mixing-DNS-pantano2002study} along the x, y, and z direction.
As a result, the cell volume used in WPTS is approximately $20$ times larger than that in the referred DNS study \cite{Tur-case-mixing-DNS-pantano2002study}.
With the coarse mesh in WPTU, the corresponding time step will increase as well by $2.45$ times.
Furthermore, the initial TKE in WPTS, $\rho E_t$, is taken as zero in the whole domain, and correspondingly no stochastic particles exist at the beginning of the simulation.
Besides, the reference mass of one stochastic particle is $2\times10^{-4} \Omega$. The CFL number is taken as 0.3, and $C_0$ is taken as 0.5. In this case, the time is normalized by $t_{\infty} = \delta_{\theta 0} / \Delta U$, namely, $t_{N} = t / t_{\infty} = t \Delta U / \delta_{\theta 0}$.

The simulation is conducted from $t_{N}=0$ to $t_{N}=1400$.
First of all, the instantaneous distribution of vorticity at different times is shown in Figure \ref{Fig-vor-wp}.
At the early stage, the flow structure develops from the initial condition and perturbation, and thus the vorticity mainly exists at the center zone (around $y=0$).
In the flow evolution, the turbulence zone becomes wider, and more flow structures at different scales with inhomogeneous distribution in the domain are generated.

\begin{figure}[htbp]
	\centering
	\subfigure{
		\includegraphics[height=1.1cm]{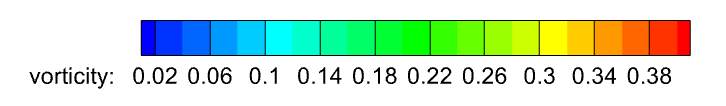}
	}
	\\
	\subfigure{
		\includegraphics[height=4.3cm]{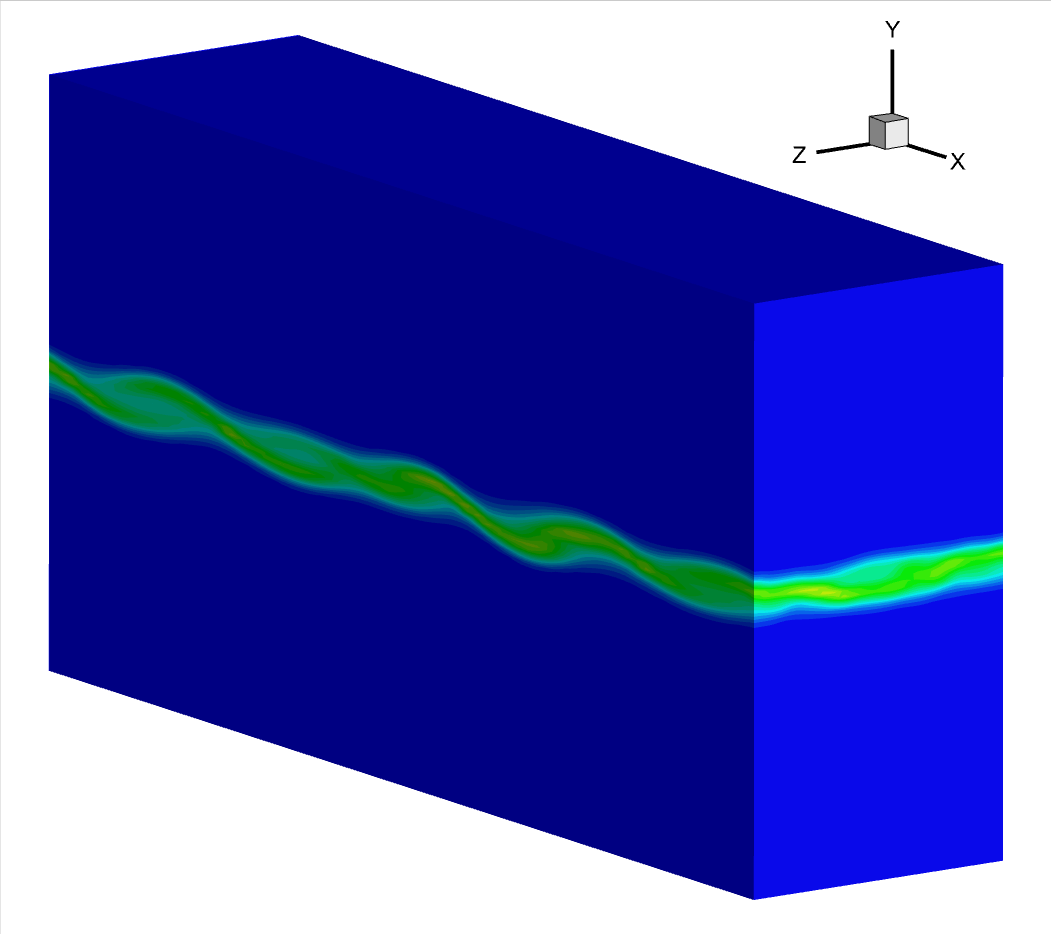}
	}
	\quad
	\subfigure{
		\includegraphics[height=4.3cm]{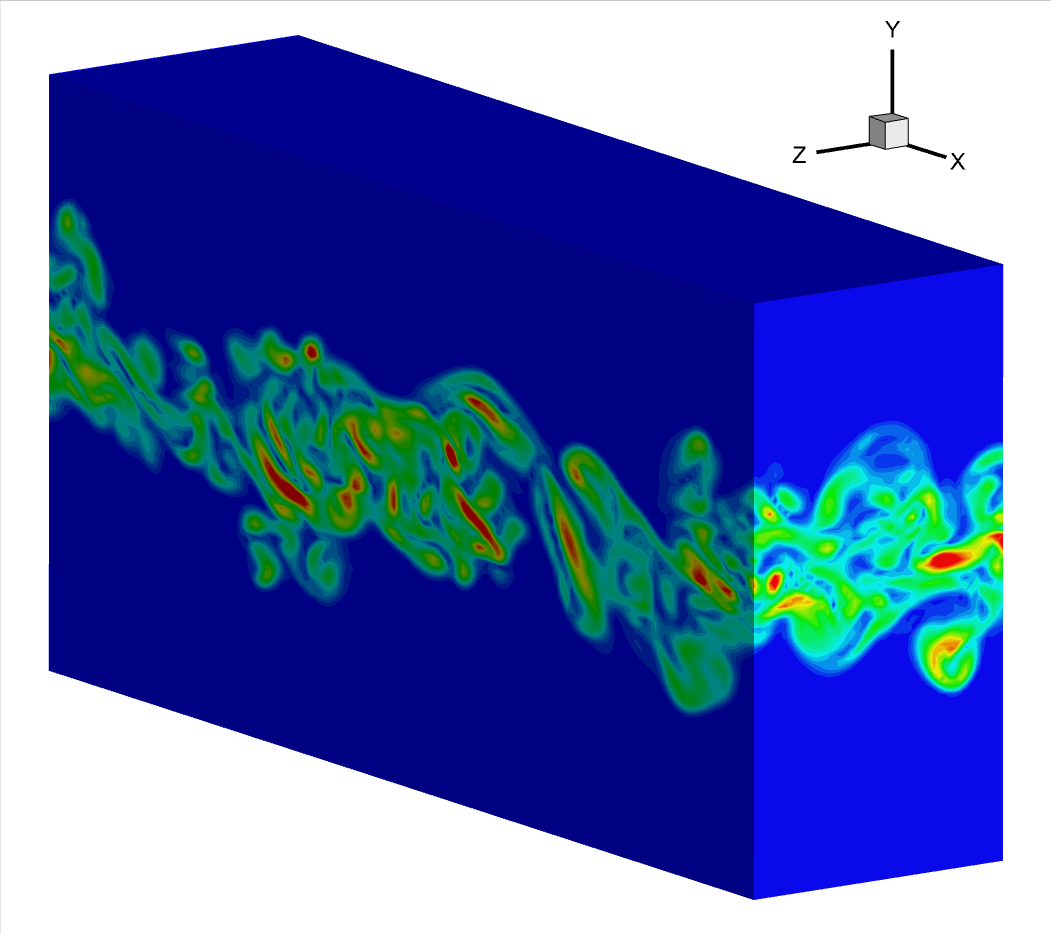}
	}
	\quad
	\subfigure{
		\includegraphics[height=4.3cm]{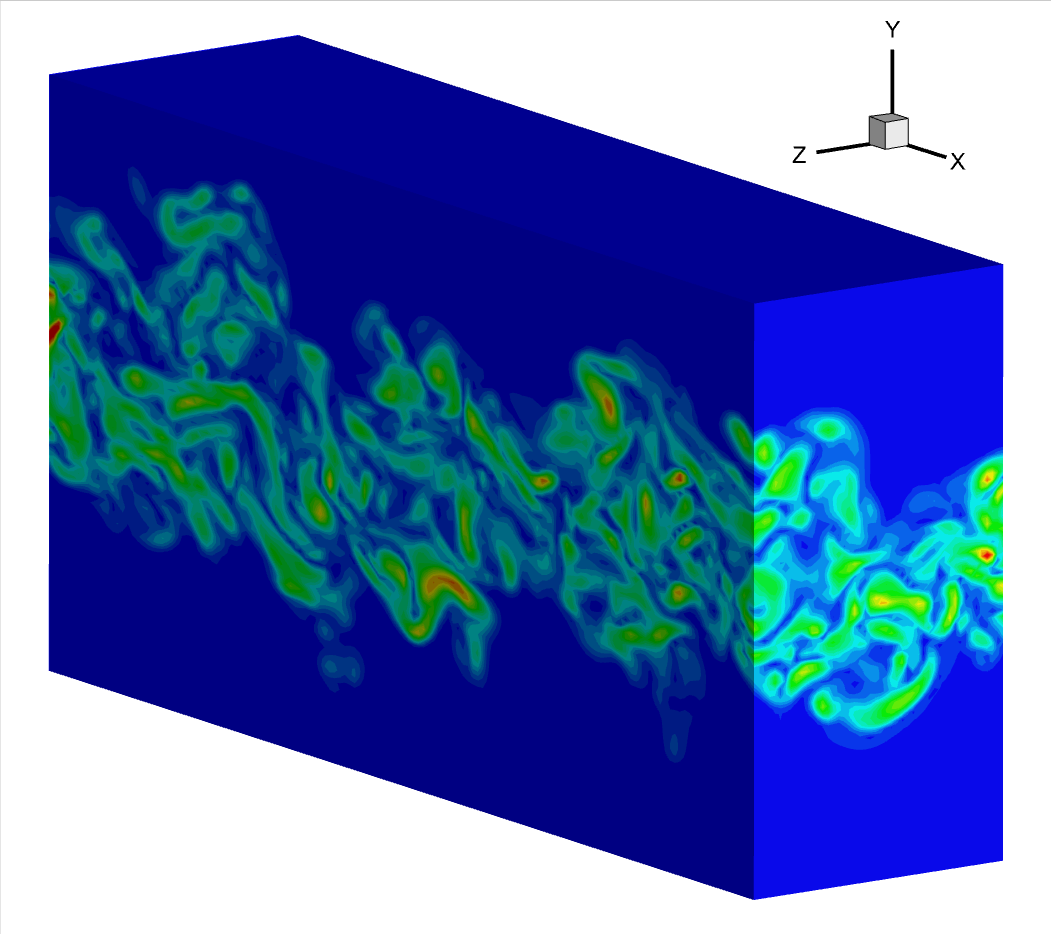}
	}
	\caption{The magnitude of vorticity  by WPTS at different times: $t_{N} = 400, 800, 1200$.}
	\label{Fig-vor-wp}
\end{figure}

To facilitate quantitative comparison with the DNS result, Figure \ref{Fig-thickness-wp} presents the time evolution of momentum thickness $\delta_{\theta}$ obtained by the WPTS method. Previous studies on compressible mixing layer turbulence have shown that, in the fully developed turbulent regime, the momentum thickness $\delta_{\theta}$ increases linearly with time, resulting in a constant slope $k$.
This constant slope $k$ is one of the typical features of the self-similarity solution, indicating the fully developed state of the mixing layer turbulence flow.
As shown in Figure \ref{Fig-thickness-wp} the linear-development feature is captured after around $t_{N}=1000$ by WPTS. Particularly, the slope $k$ obtained by WPTS is 0.0095. It agrees well with the reference value $k=0.0108$ by DNS in \cite{Tur-case-mixing-DNS-pantano2002study}, despite a slight deviation.

\begin{figure}[htbp]
	\centering
	\includegraphics[height=6.0cm]{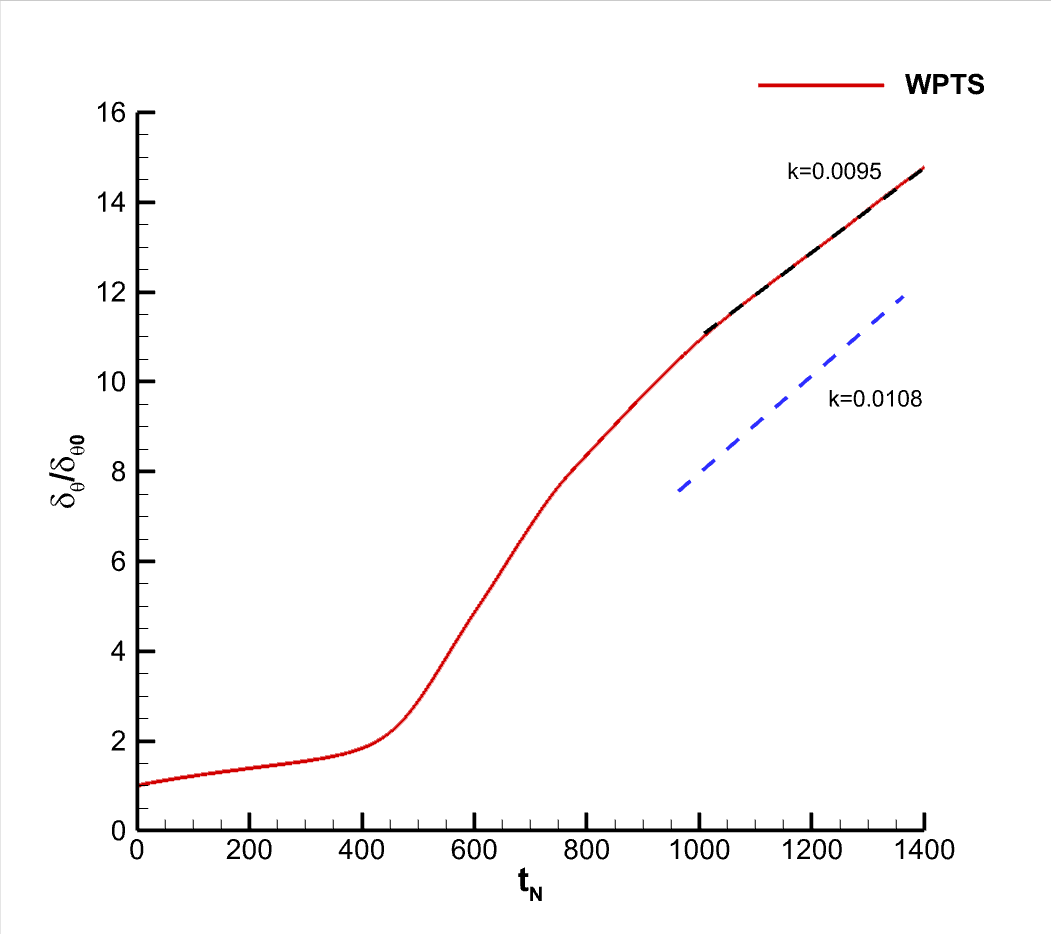}
	\caption{The development of momentum thickness $\delta_{\theta}$ by WPTS, indicating the slope $k=0.0095$ in the self-similarity solution which is presented by the black dashed line. The slope $k=0.0108$ shown by a blue dashed line is from the DNS study of Pantato.}
	\label{Fig-thickness-wp}
\end{figure}

To further validate the WPTS, the Reynolds-stress associated terms are presented in Figure \ref{Fig-Rij-wp}.
As important turbulence statistic variables, the Reynolds stress terms $R_{ij}$ at the fully-developed stage in mixing layer turbulence, have been widely studied by both experiment measurement and numerical simulation.
Particularly, the $R_{ij}$ should give the nearly same distribution in the fully-developed state, which is the key feature in the so-called self-similarity solution.
Figure \ref{Fig-Rij-wp} gives the normalized variables $\sqrt{R_{xx}}/\Delta U$, $\sqrt{R_{yy}}/\Delta U$, and $\sqrt{R_{xy}}/\Delta U$ by WPTS under different times, namely $t_N=1100, 1200, 1300$.
In Figure \ref{Fig-Rij-wp}, the $y_N = y / \delta_{\omega}$ indicates the transverse position normalized by the vorticity thickness $\delta_{\omega}$, where the vorticity thickness can be obtained by $\delta_{\omega} = \Delta U /  | \partial \widetilde{U} / \partial y | _{max}$ and is estimated by $\Delta U /  | \partial \widetilde{U} / \partial y |_{y=0}$ in this work.

The results presented in Figure \ref{Fig-Rij-wp} demonstrate that, while minor discrepancies exist between the solutions obtained by the WPTS method at three different times, the overall agreement with the reference solutions from previous DNS and experimental studies is satisfactory \cite{Tur-case-mixing-EXP-elliott1990compressibility, Tur-case-mixing-DNS-pantano2002study}. It is important to note that the DNS reference solution is obtained by averaging the results over several time instances \cite{Tur-case-mixing-DNS-pantano2002study}. Consequently, the $\sqrt{R_{ij}}/\Delta U$ obtained by WPTS at $t_N=1100, 1200, 1300$ are averaged and shown in Figure \ref{Fig-Rij-wp}.
This averaging yields excellent agreement between the WPTS results (the three Reynolds-stress associated terms) and the DNS reference value, both in terms of profile width and peak magnitude. Therefore, based on the results presented in Figure \ref{Fig-thickness-wp} and Figure \ref{Fig-Rij-wp}, we conclude that the proposed WPTS method accurately captures the typical features of the fully developed turbulence in the CML case, providing strong evidence for its reliability and potential in turbulence simulation under a coarse mesh resolution.

Furthermore, it is emphasized that we have conducted gigantic amount of computations to this case.
As a deterministic-stochastic method, the WPTS produces slightly different results in each computation.
With identical parameters, the best and worst results in comparison with the reference solutions are provided in Appendix D. All computations give consistent results in the compressible mixing layer studies.

\begin{figure}[htbp]
	\centering
	\subfigure{
		\includegraphics[height=4.2cm]{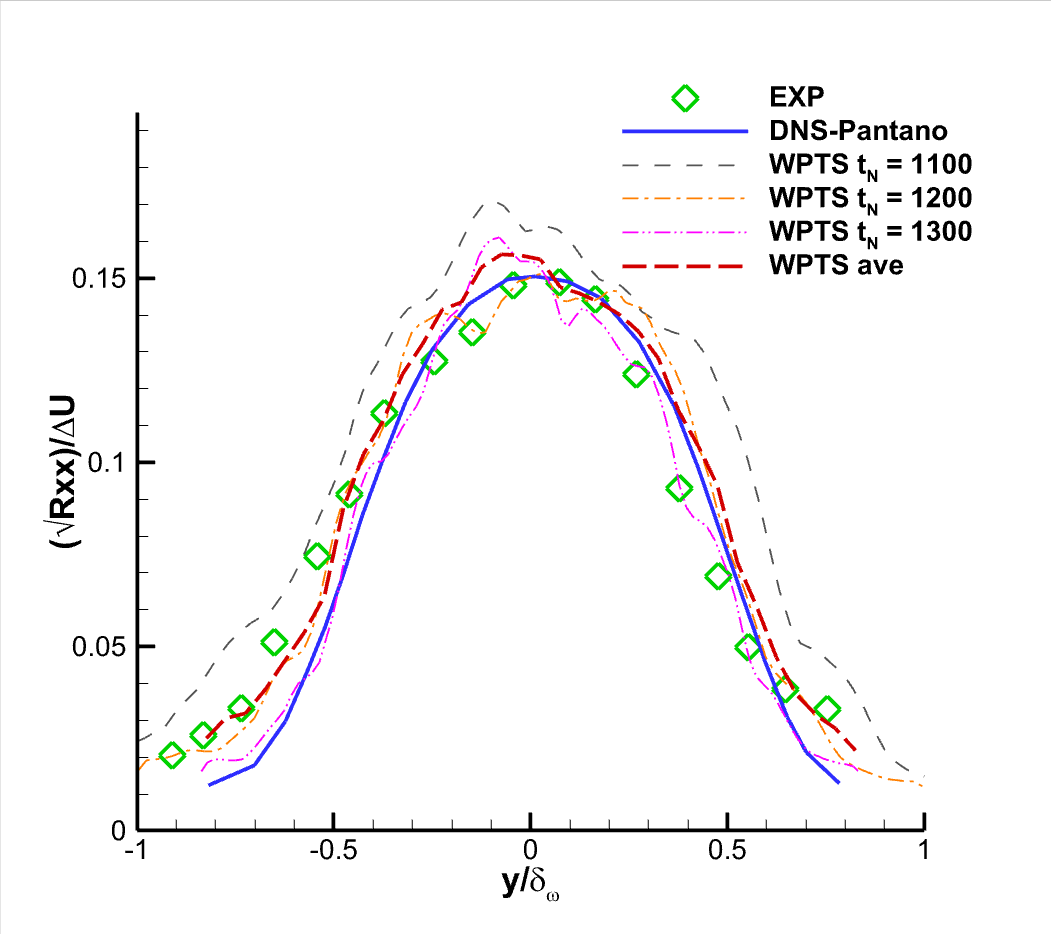}
	}
	\quad
	\subfigure{
		\includegraphics[height=4.2cm]{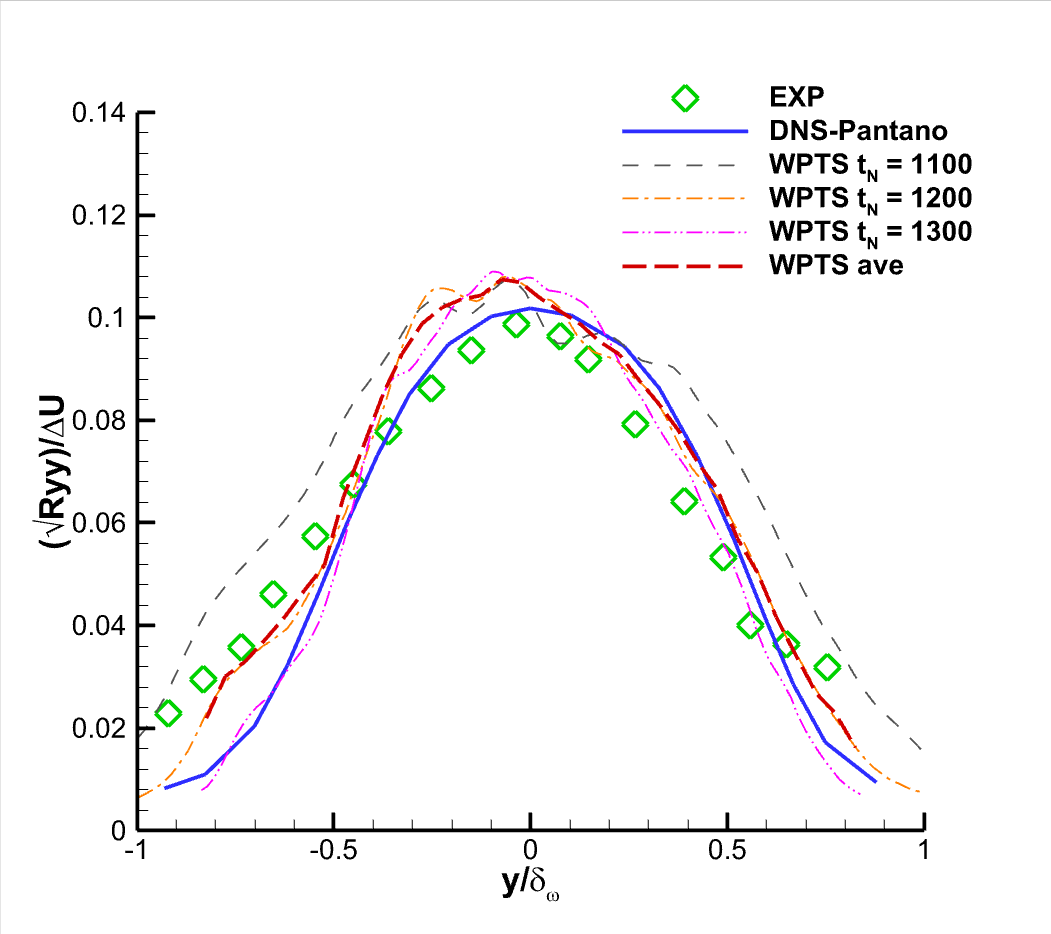}
	}
	\quad
	\subfigure{
		\includegraphics[height=4.2cm]{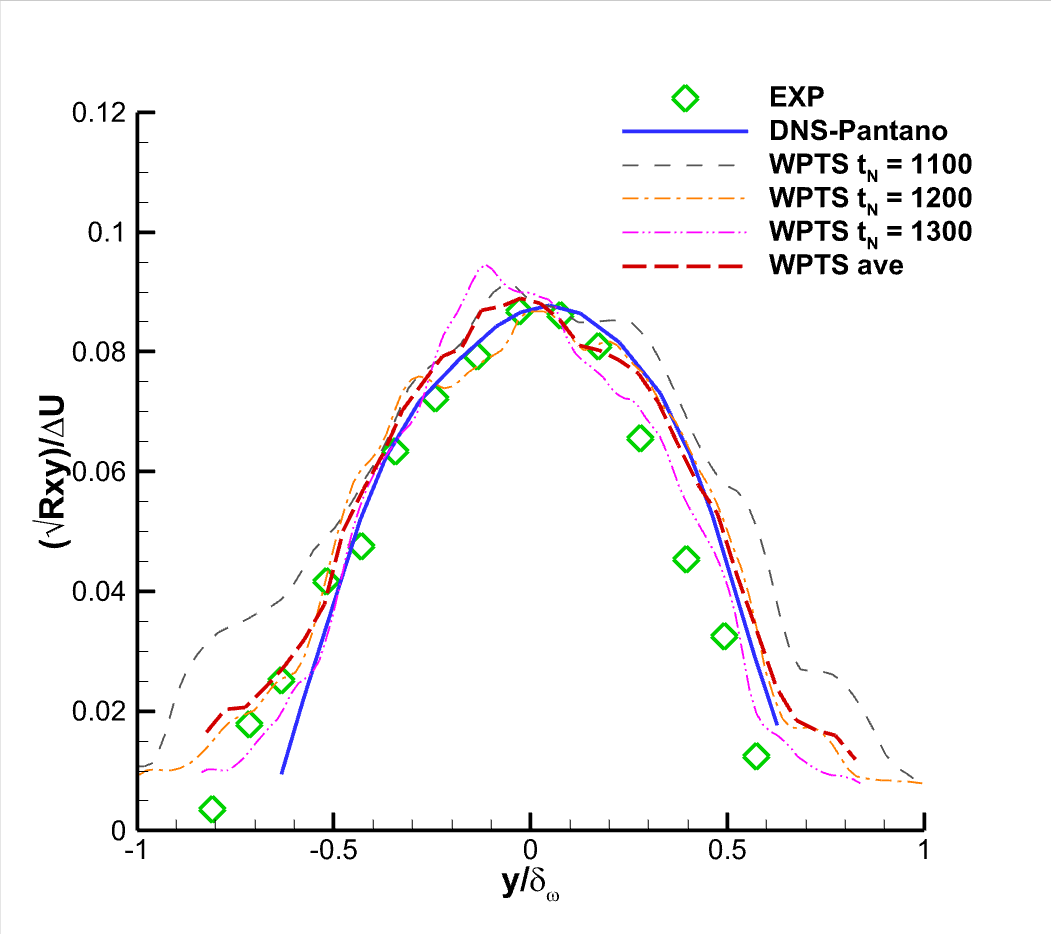}
	}
	\caption{The Reynolds stress terms $\sqrt{R_{xx}}/\Delta U$, $\sqrt{R_{yy}}/\Delta U$ and $\sqrt{R_{xy}}/\Delta U$ by WPTS at different times $t_{N} = 1100, 1200, 1300,$ and the averaged counterpart. The referent results by DNS and experiments are from \cite{Tur-case-mixing-DNS-pantano2002study} and \cite{Tur-case-mixing-EXP-elliott1990compressibility}, respectively.}
	\label{Fig-Rij-wp}
\end{figure}

As a comparison, the direct NS solution under the same coarse mesh is presented as well by using the fifth-order GKS in Figure \ref{Fig-Rij-iLES}.
It is worth noting that all the settings in the numerical simulation are the same except the solver.
Here the GKS is used for the purely NS solution with a coarse mesh, which can be used as a DNS solver once the mesh resolution is fine enough.
As shown in the figure, the distributions of $\sqrt{R_{ij}}/\Delta U$ by GKS deviate from the reference solutions, such as the width and the values at the peak zone.
It indicates that the modeling in WPTS improves the solution for the turbulence simulation under unresolved mesh resolution.

\begin{figure}[htbp]
	\centering
	\subfigure{
		\includegraphics[height=4.2cm]{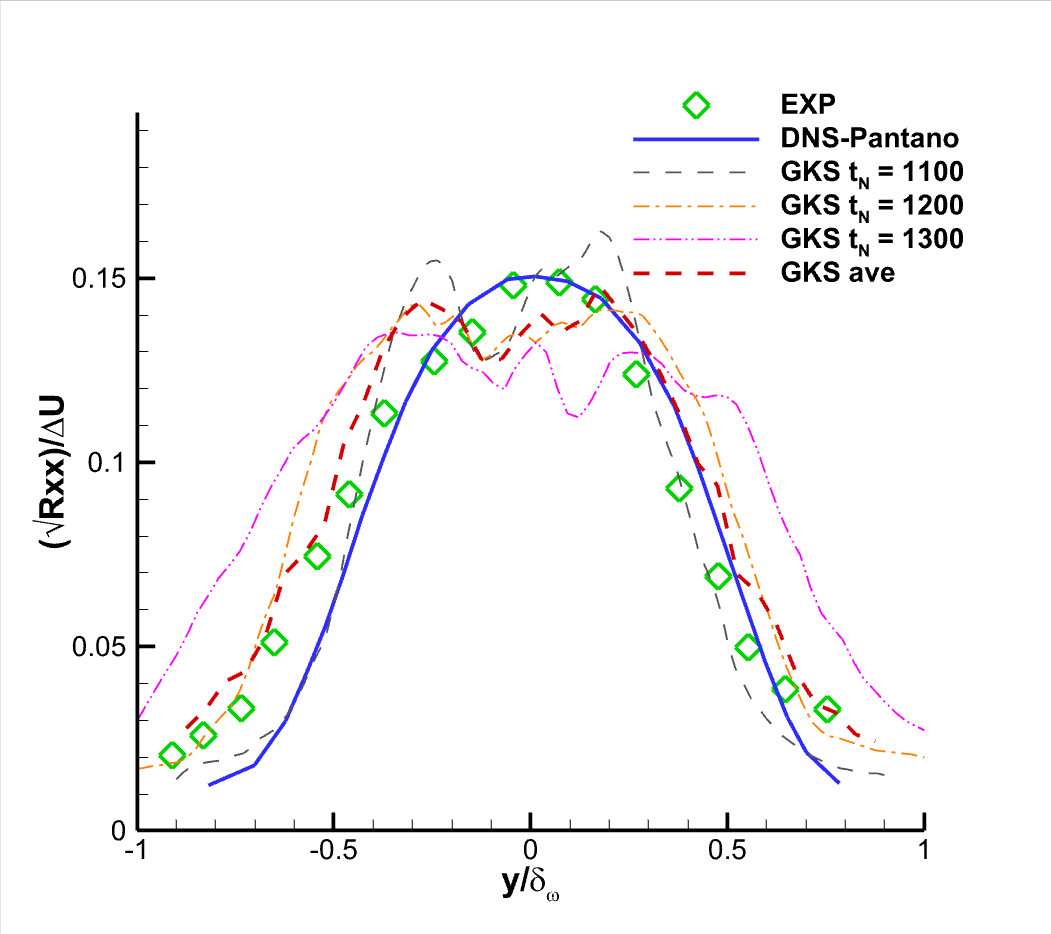}
	}
	\quad
	\subfigure{
		\includegraphics[height=4.2cm]{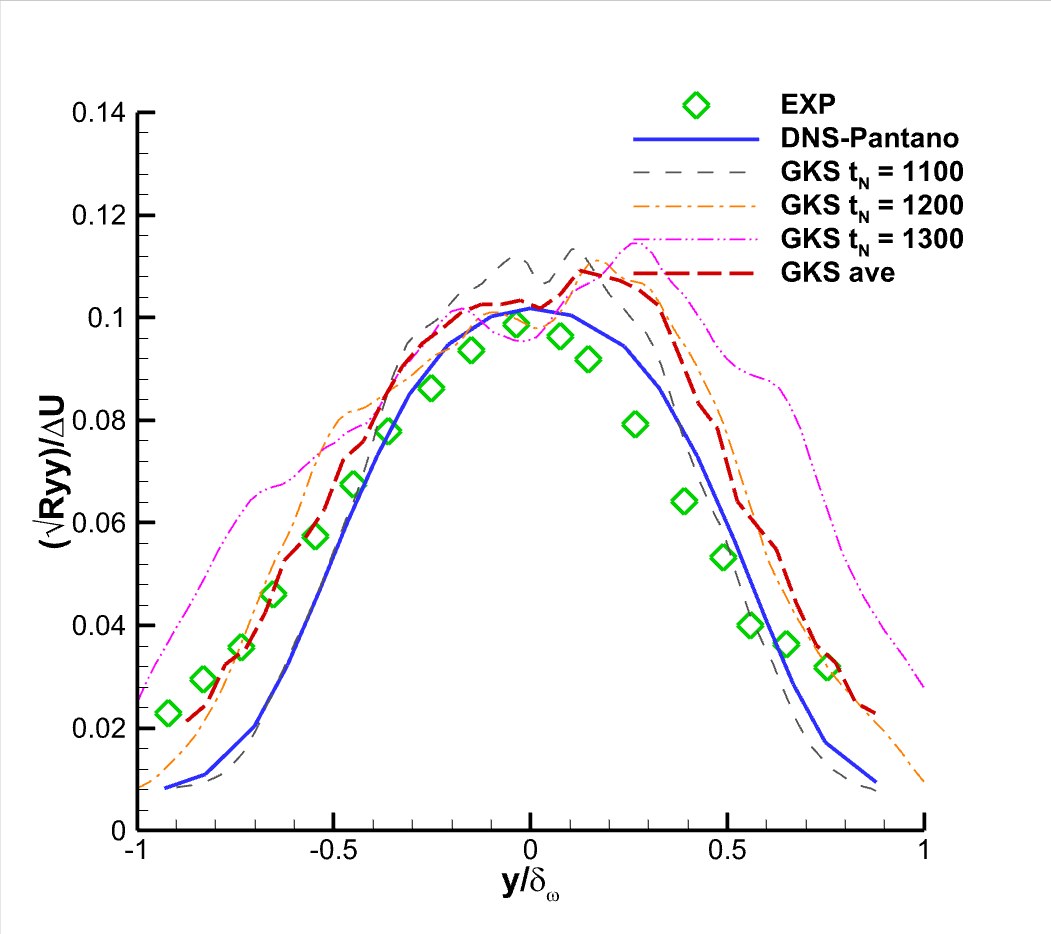}
	}
	\quad
	\subfigure{
		\includegraphics[height=4.2cm]{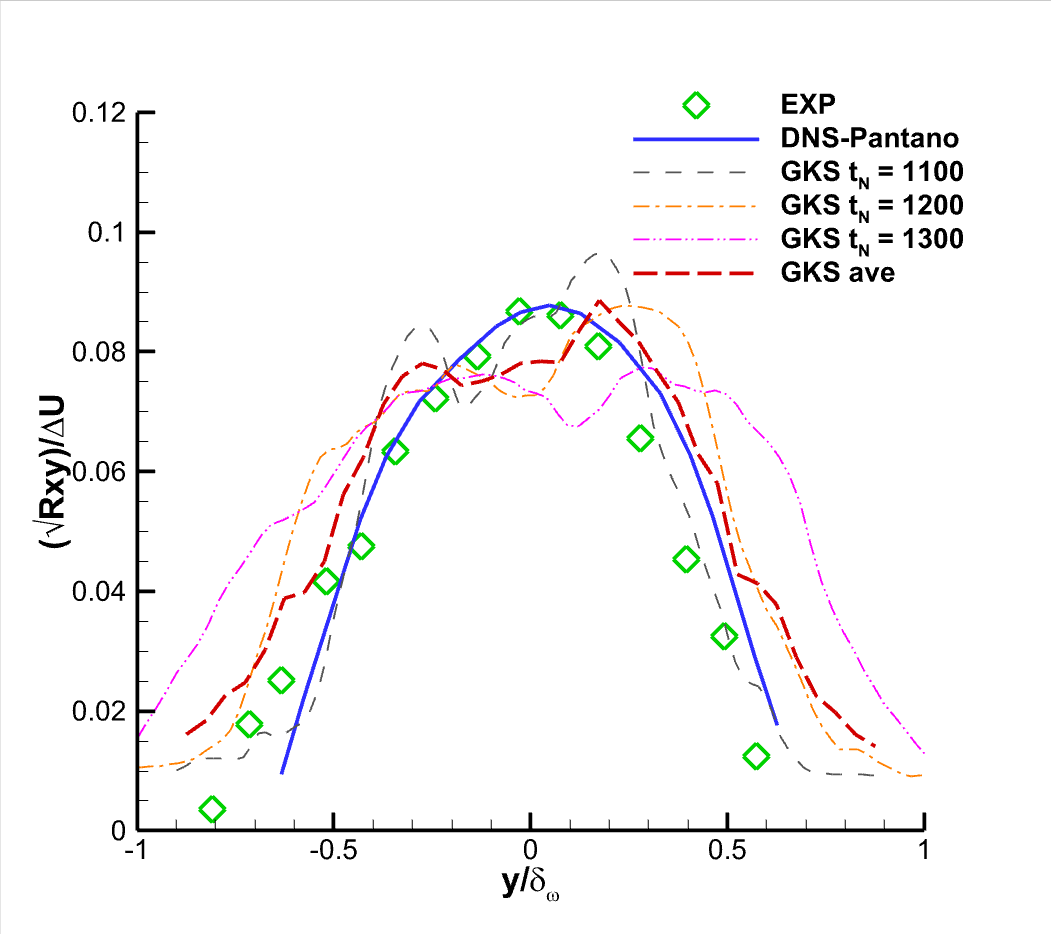}
	}
	\caption{The Reynolds stress terms $\sqrt{R_{xx}}/\Delta U$, $\sqrt{R_{yy}}/\Delta U$ and $\sqrt{R_{xy}}/\Delta U$ by GKS at different times $t_{N} = 1100, 1200, 1300,$ and the averaged counterpart, indicating the NS solution under such an un-resolved grid.
	The referent results by DNS and experiments are from \cite{Tur-case-mixing-DNS-pantano2002study} and \cite{Tur-case-mixing-EXP-elliott1990compressibility}, respectively.}
	\label{Fig-Rij-iLES}
\end{figure}

Another widely employed turbulent simulation method based on the coarse mesh is the LES. Here the GKS is extended as a LES solver, where the physical viscosity is replaced by the turbulence counterpart determined by the SM model, where the coefficient in the SM model is taken as $C_s^2=0.015$.
The results of instantaneous and averaged Reynolds stresses are presented in Figure \ref{Fig-Rij-LES}.
The results indicate that the LES with SM cannot well predict the $\sqrt{R_{ij}}/\Delta U$ under such a coarse mesh. The averaged values around the peak zone are obviously higher than the reference solutions. In other words, if we consider the direct GKS solution with physical viscosity as implicit LES (iLES), the GKS solutions show that the iLES results are better than those from the LES modeling.

\begin{figure}[htbp]
	\centering
	\subfigure{
		\includegraphics[height=4.2cm]{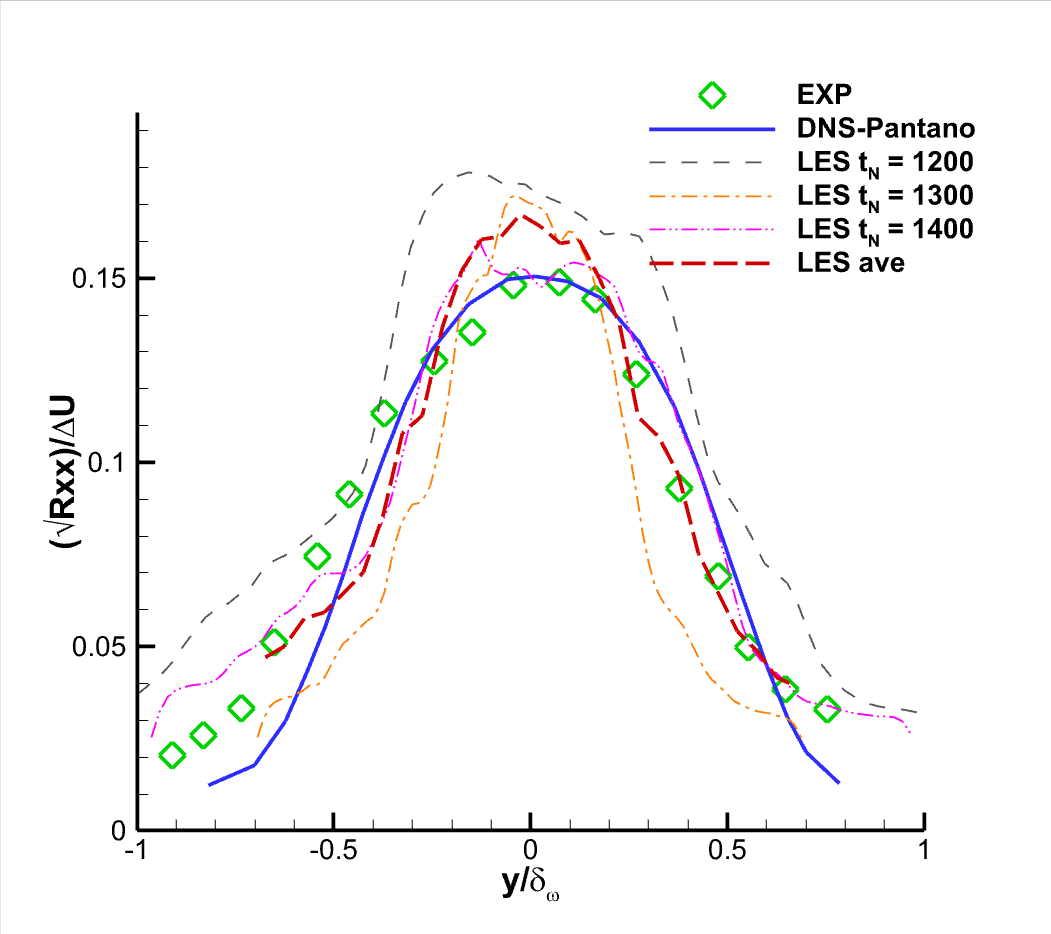}
	}
	\quad
	\subfigure{
		\includegraphics[height=4.2cm]{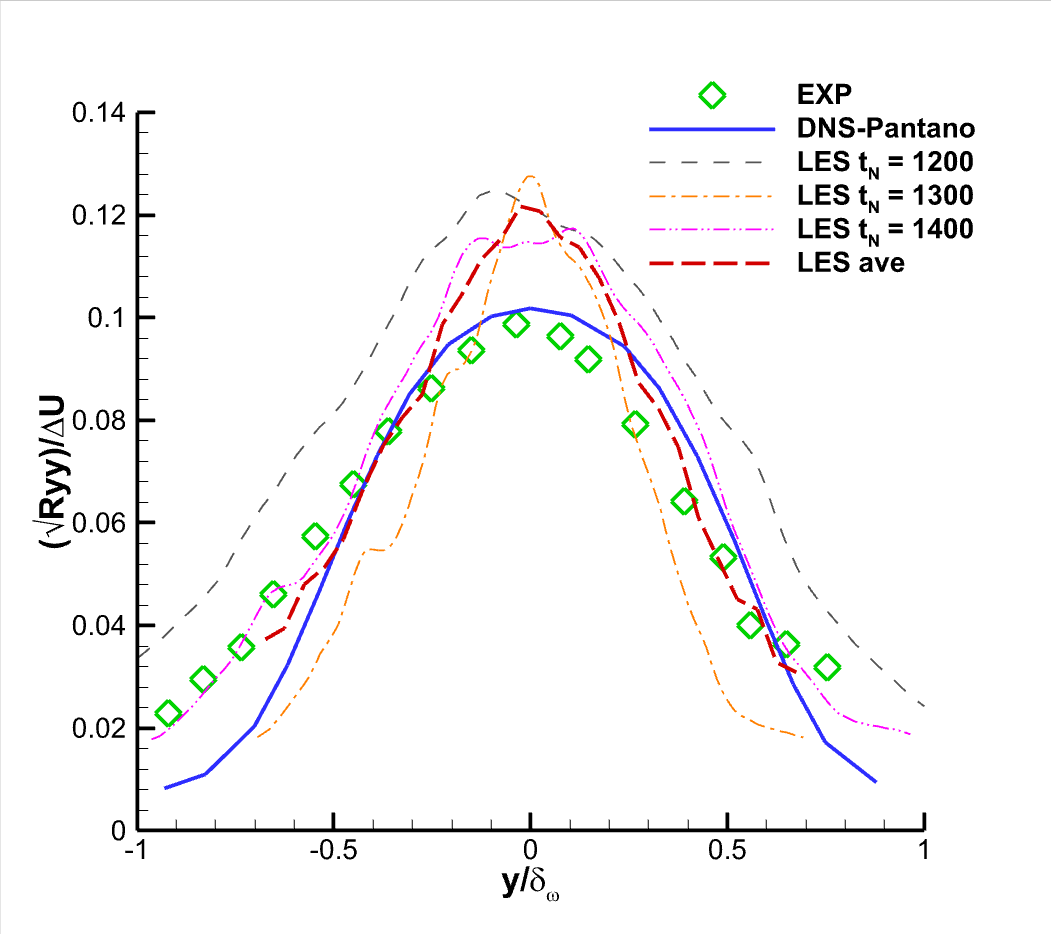}
	}
	\quad
	\subfigure{
		\includegraphics[height=4.2cm]{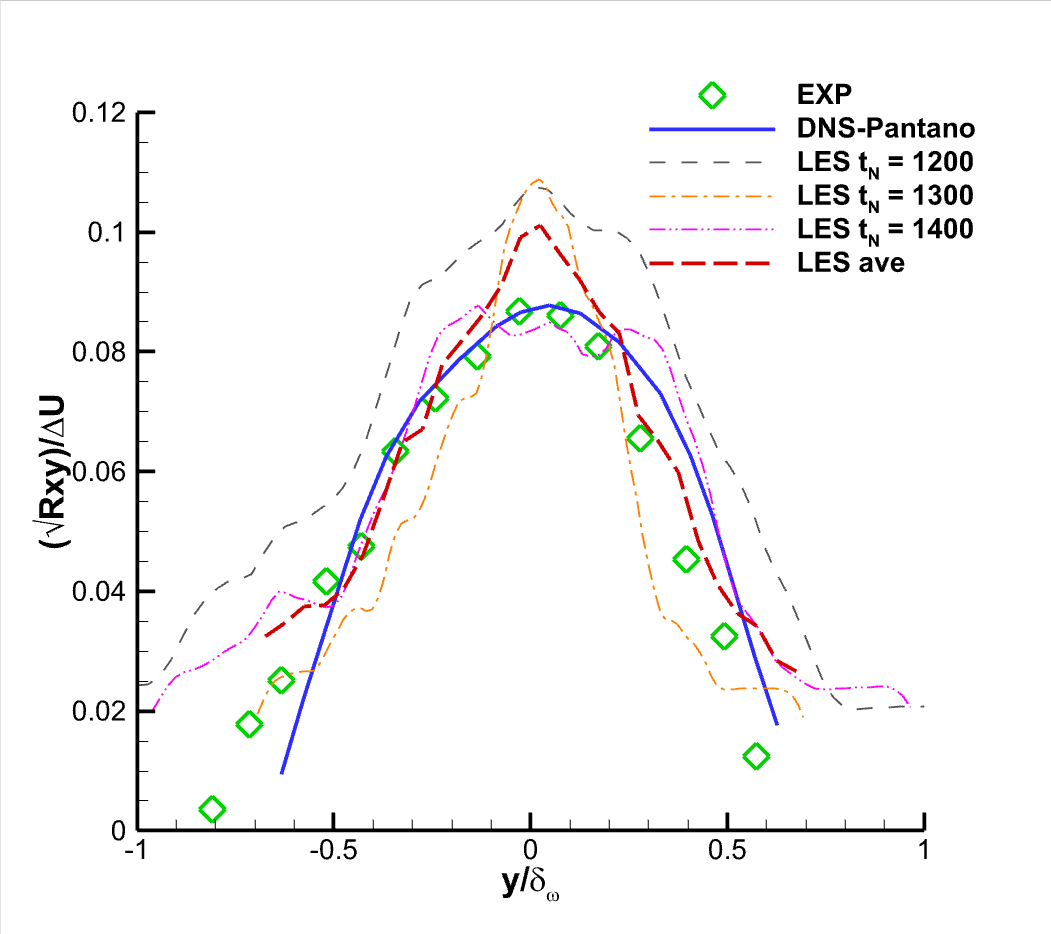}
	}
	\caption{The Reynolds stress terms $\sqrt{R_{xx}}/\Delta U$, $\sqrt{R_{yy}}/\Delta U$ and $\sqrt{R_{xy}}/\Delta U$ by LES based on GKS solver at different times $t_{N} = 1200, 1300, 1400,$ and the averaged counterpart. The SM model is employed in LES and $C_s^2=0.015$. The referent results by DNS and experiments are from \cite{Tur-case-mixing-DNS-pantano2002study} and \cite{Tur-case-mixing-EXP-elliott1990compressibility}, respectively.}
	\label{Fig-Rij-LES}
\end{figure}

The instantaneous flow field from WPTS, especially the distribution of stochastic particles, will be presented and analyzed here.
Firstly, Figure \ref{Fig-sij-tke} gives the instantaneous flow field at plane $z=0$ and time $t_{N} = 1200$. Particularly, Figure \ref{Fig-sij-tke}(a) and Figure \ref{Fig-sij-tke}(b) are  the distributions of the streamwise velocity $U$ and the velocity strain $|\vec{S}|$ in Eq.\eqref{taumac}, respectively.
Besides, Figure \ref{Fig-sij-tke}(c) shows the whole collision time determined by $\tau_{n} = \tau_{p} + \tau_{t}$.
The results by WPTS show that the turbulence collision time $\tau_{t}$ is much larger than the physical one $\tau_{p}$ in the region with strong turbulence intensity.
Correspondingly, the TKE is also presented in Figure \ref{Fig-sij-tke}(d), which indicates the local velocity fluctuation due to discrete moving particles over the background flow field.

\begin{figure}[htbp]
	\centering
	\subfigure[]{
		\includegraphics[height=4.2cm]{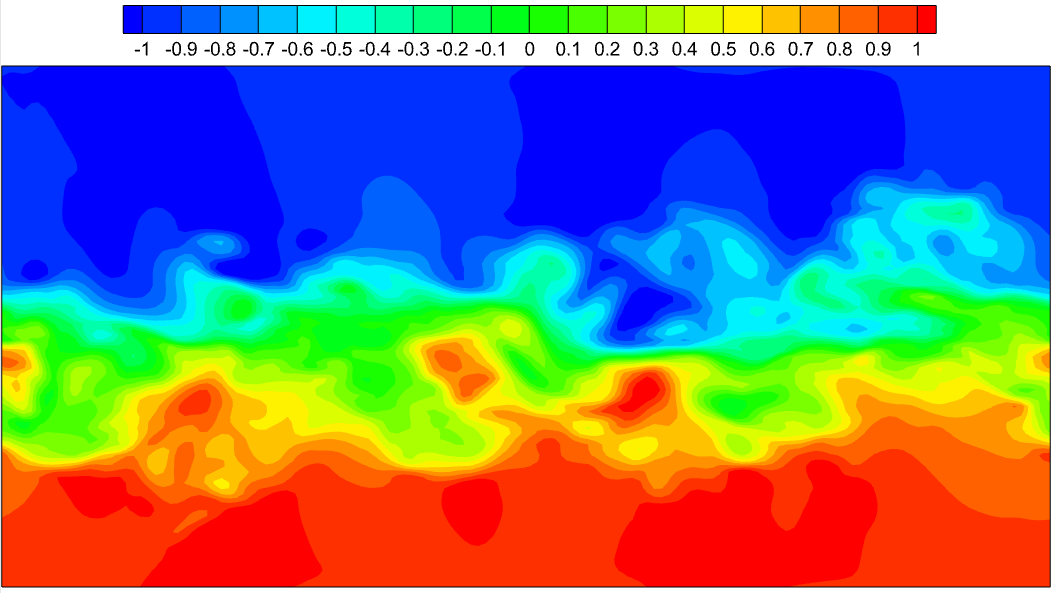}
	}
	\quad
	\subfigure[]{
		\includegraphics[height=4.2cm]{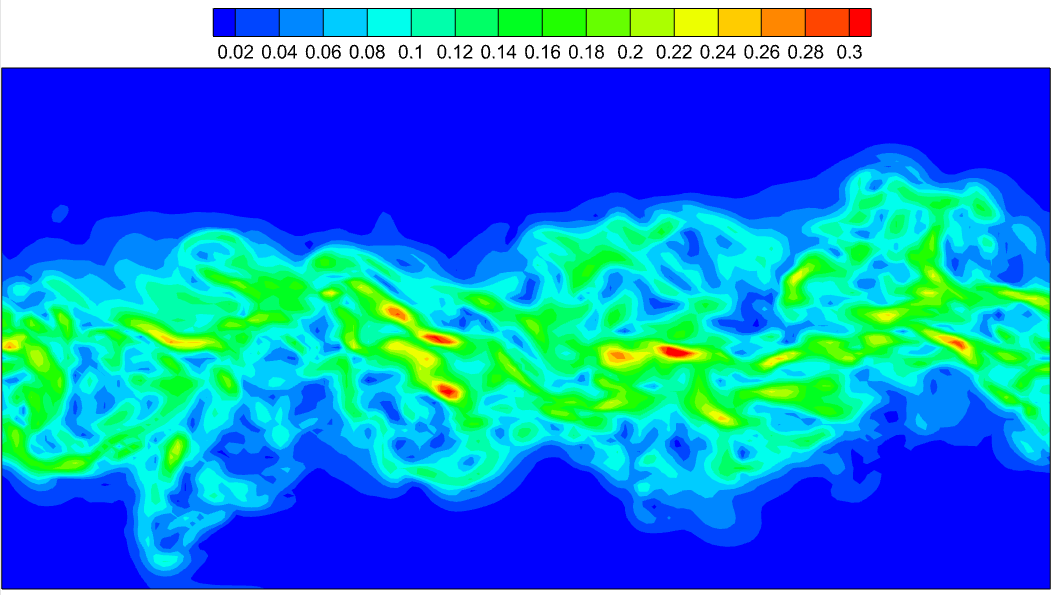}
	}
	\\
	\subfigure[]{
		\includegraphics[height=4.2cm]{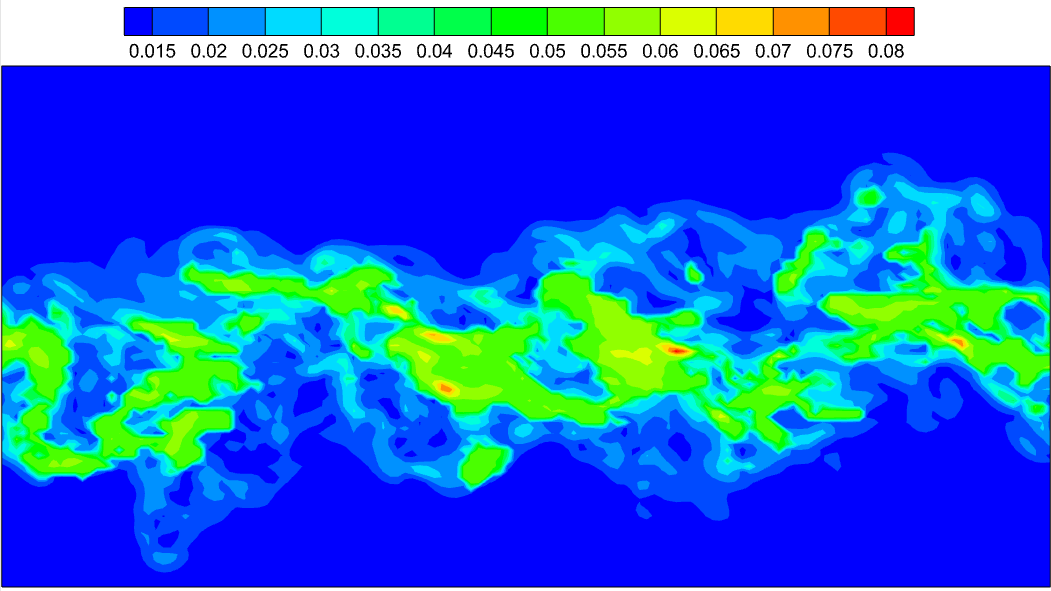}
	}
	\quad
	\subfigure[]{
		\raisebox{0.001\textwidth}{
			\includegraphics[height=4.2cm]{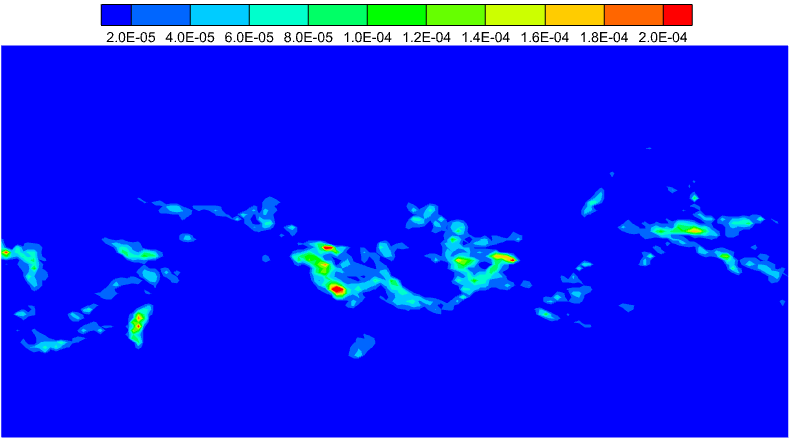}
		}		
	}
	\caption{The snapshot of flow field for plane $z=0$ at time $t_{N} = 1200$ by WPTS: (a) the streamwise velocity $U$, (b) the term of velocity strain $|\vec{S}|$ in Eq.\eqref{taumac}, (c) the collision time $\tau_{n} = \tau_{p} + \tau_{t}$, (d) the sub-cell turbulence kinetic energy $\rho E_t$ obtained by the surviving particles.}
	\label{Fig-sij-tke}
\end{figure}

\begin{figure}[htbp]
	\centering
	\subfigure[]{
		\includegraphics[height=4.2cm]{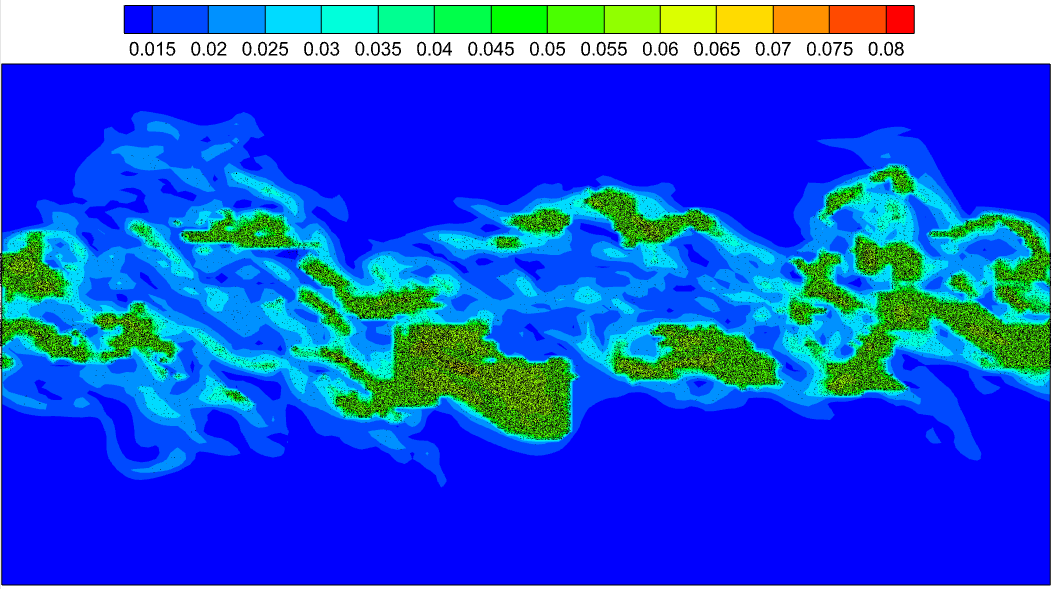}
	}
	\subfigure[]{
		\includegraphics[height=4.2cm]{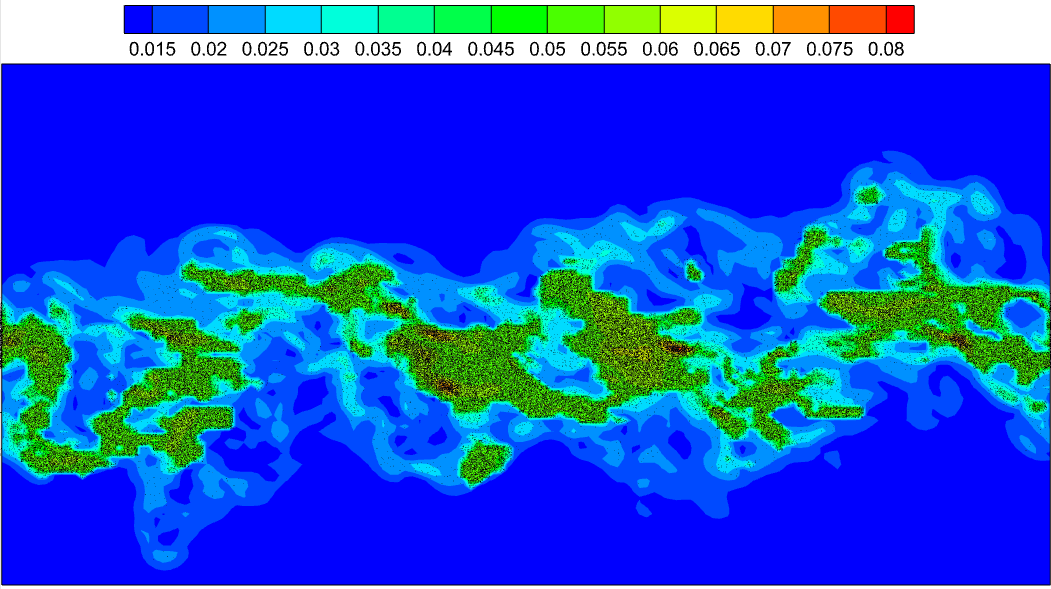}
	}
	\subfigure[]{
		\includegraphics[height=4.2cm]{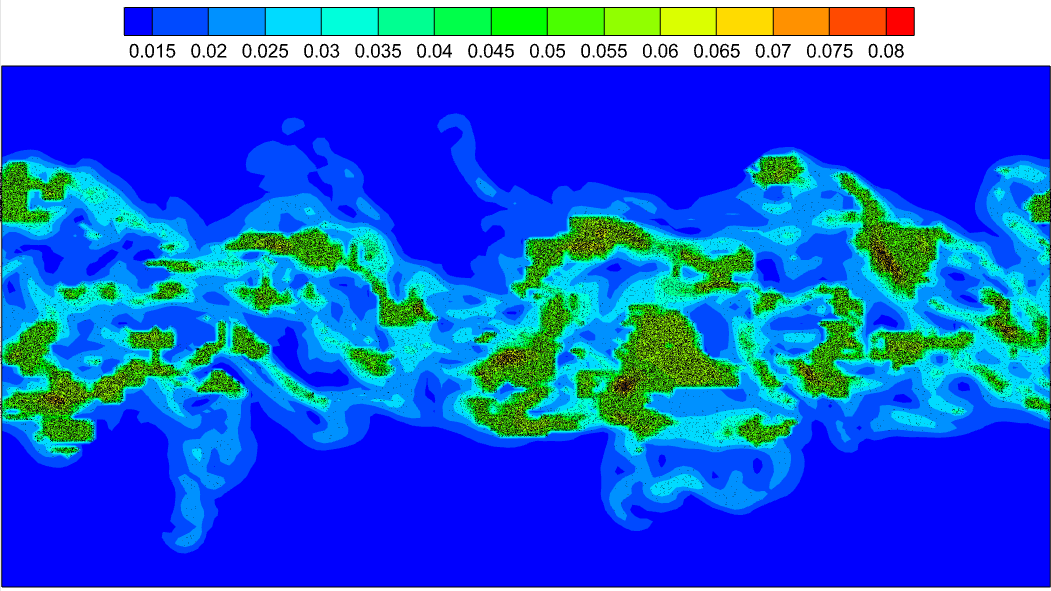}
	}	
	\caption{The scattering of stochastic particles in WPTS with the background contour of $\tau_{n}$ at time $t_{N} = 1200$. The results in (a), (b), and (c) are for different planes in the transverse direction, namely $\frac{1}{6}Z, \frac{3}{6}Z, \frac{5}{6}Z$ respectively.}
	\label{Fig-par-tau}
\end{figure}

Figure \ref{Fig-par-tau} shows the scattering of stochastic particles on three different planes in the transverse direction at $t_N = 1200$. The background contour is given by $\tau_{n}$. Since the percentage of the stochastic particles in the current WPTS framework is mainly determined by $\tau_n$, the stochastic particles distributions have been mainly concentrated in the regions with high $\tau_n$.
Figure \ref{Fig-par-tau}(b) shows the results on the same plane and at the same time as Figure \ref{Fig-sij-tke}. The results indicate that the particle component is highly related to $\vec{S}$ value, due to the adoption of $\tau_t$ model in Eq.\eqref{tautur}.
Since the sub-cell TKE in WPTS is directly obtained by counting the surviving stochastic particles, as given in Appendix B, the TKE value is also overwhelmingly higher in the region with large $\tau_t$.
Overall, the stochastic particles in WPTS models the sub-grid turbulent flow dynamics, and the wave provides the cell-resolved large-scale flow evolution.
The number of particles depends on the cell resolution, such as the local time step, and discrete fluid elements collision time.
Particularly, the non-equilibrium transport of particle penetration is included in the turbulence modeling.
The simulations results confirm that the current framework deserves further study.

\section{Conclusion}
In this paper, a new framework called WPTS is proposed for turbulence simulation under a coarse mesh.
Different from the widely-used RANS and LES approaches, WPTS decomposes the fluid field into wave and particle components, where the wave component mainly models the resolved flow structures under the employed grid, and the particle component mainly stands for the evolution of the discrete fluid elements, which directly models the unresolved turbulent flow. In other words, the WPTS is mainly a deterministic-stochastic method.
The decomposition of wave and particle is determined by the turbulence collision time $\tau_{t}$, which has the similar consideration as Smagorinsky model, but with different physical understanding.
The current wave-particle framework will automatically adjust the wave-particle decomposition, and the stochastic particles mainly appear and play roles in the flow region with strong turbulence intensity.
For a well-resolved flow structure, the particles in WPTS will disappear and the WPTS will recover GKS for the laminar flow simulation, like a DNS method.
The wave-particle adaptation appears in the coarse mesh condition and effectively saves computational cost for turbulent study.
The multiscale nonequilibrium transport mechanism is built intrinsically in the WPTS. There is no a clear distinction between the laminar and turbulent flows anymore.

The typical turbulence case, namely temporal compressible mixing layer, is employed to validate the performance of WPTS. The results show good agreement with the DNS and experimental measurement with the capturing of typical turbulence features, such as the development of momentum thickness, Reynolds stress terms, etc.
The results by purely iLES and LES Navier-Stokes solvers, like the GKS-based methods, are also presented for comparison.
Obvious improvements from WPTS are observed and provide evidence for the reliability and potential of WPTS modeling for turbulent flow.

In future, the further developments of WPTS can consider the following aspects.
Firstly,  the more delicate turbulence collision time should be studied and developed, which is the key factor for the wave-particle decomposition.
There are much more space than the simple adaptation of  Smagorinsky-type model here.
Secondly, besides the pressure gradient, the forcing on the particle can be further refined.
For example, the current particle has only the kinetic energy to recover the TKE in turbulent flow. Theoretically, the particle here represents broken hydrodynamic fluid element, which could have rotational energy as well, such as the moving vortices. The element's rotational motion can be modeled as internal degree of freedom of the particle. Therefore, the forcing on the rotating macro-particle should be carefully modeled.
Thirdly, for the wall turbulence, the particles' evolution in the near wall region should be studied with the consideration of the effect from the wall, especially with
the unresolved coarse mesh. The corresponding WPTS with wall model has to be developed.
Finally, the current non-equilibrium transport from the particle has similar mechanism as the molecular motion in the rarefied flow. Since the corresponding UGKWP method for the rarefied flow has been developed and validated in a gigantic amount of engineering applications. With the combination of WPTS and UGKWP, it is expected a unified flow solver from the turbulent flow to the rarefied one can be constructed, which may play an important role in the studies of turbulent flow transition around a space vehicle in the continuum-rarefied flow environment.

\section{Acknowledgements}
The authors would like to thank Dr. Yajun Zhu and Dr. Chang Liu for helpful discussion. The current research is supported by National Key R\&D Program of China (Grant Nos. 2022YFA1004500), National Science Foundation of China (12172316, 92371107), and Hong Kong research grant council (16301222, 16208324).

\setcounter{equation}{0}
\renewcommand\theequation{A.\arabic{equation}}
\section*{Appendix A: The definition of total energy}
For the given $g$ in Eq.\eqref{geq},
the relation between macroscopic variables $\vec{W}$ and the distribution function $g$ can be obtained by taking moments with $\vec{\psi}$
\begin{gather*}
\left\langle \vec{\psi} \right\rangle = \left[\rho,~\rho \vec{U},  ~\frac{1}{2}\rho\vec{U}^2 + \frac{K+3}{2} \frac{\rho}{2\lambda}\right]^T.
\end{gather*}
Note that the above $\lambda$ includes both thermal and turbulent kinetic energy. In other words, we have
\begin{align*}
\rho E &\overset{def}{=} \frac{1}{2}\rho\vec{U}^2 + \frac{K+3}{2} \frac{\rho}{2\lambda} \\
&= \frac{1}{2}\rho\vec{U}^2 + \frac{K+3}{2} \rho R T \\
&= \frac{1}{2}\rho\vec{U}^2 + \frac{K}{2} \rho R T_{thermal}
+ \frac{3}{2} \rho R T_{thermal} + \frac{3}{2} \rho \Theta_t\\
&= \frac{1}{2}\rho\vec{U}^2 + \frac{K+3}{2} \rho R T_{thermal}
+ \frac{3}{2} \rho \Theta_t,
\end{align*}
where the last term on the right-hand side (RHS) of $\rho E$, $\frac{3}{2} \rho \Theta_t$, also denoted as $\rho E_t$, is the TKE.

It is noted  that in the current wave-particle method, the $\rho E_t$ can be directly obtained by counting the surviving particles (referring to Appendix B). Therefore if needed, the thermal temperature $T_{thermal}$ can be obtained based on the $\rho E$ and $\rho E_t$.

\setcounter{equation}{0}
\renewcommand\theequation{B.\arabic{equation}}
\section*{Appendix B: The evaluation of TKE from the surviving particles}

With the TKE definition
\begin{gather*}
	\rho E_t \overset{def}{=}\frac{1}{2}\rho \left(u^{'}u^{'}+v^{'}v^{'} + w^{'}w^{'}\right),
\end{gather*}
where $\vec{u}^{'}=\left[u^{'}, v^{'}, w^{'}\right]^T$ is the fluctuation velocity.
In each cell, the fluctuation velocity $\vec{u}^{'}$ or $\delta \vec{u}_p$ from a particle is expressed as
\begin{equation*}
\vec{u}^{'} = \vec{u}_p - \vec{U},
\end{equation*}
where $\vec{u}_p$ is the velocity of the stochastic particle and $\vec{U}$ is the cell-averaged macroscopic velocity.
As a result, the cell-averaged TKE value, carried by surviving particles, can be evaluated directly by counting as below
\begin{equation}
\rho E_{t} =\sum_{k} \frac{m_p}{V} \frac{1}{2}\vec{u}^{'}_{k} \cdot \vec{u}^{'}_{k}.
\end{equation}

\setcounter{equation}{0}
\renewcommand\theequation{C.\arabic{equation}}

\section*{Appendix C: Sampling particles based on the TKE}

Overall, for the particle sampling in the wave-particle method, we need to know the density $\rho^*$ and the TKE $\rho E_t^*$ carried by the newly sampled stochastic particles.
With the modeling of TKE being carried by newly sampled particles with normal distribution and isotropic property,
the particles' velocity $\vec{u}_p$ can be determined by
\begin{gather}\label{veldel-sampling}
\delta \vec{u}_p = \mathcal{D}_{N} \left[\rho E_t^*, \rho^*\right] \overset{def}{=} \vec{\omega} \sqrt{\frac{2}{3} \frac{\rho E_t^*}{\rho^*}}, ~~ \vec{\omega} = [\omega_1, \omega_2, \omega_3]^T, ~~ \omega_i \in \mathcal{N}(0,1),
\end{gather}
\begin{gather}\label{vel-sampling}
	\vec{u}_p = \delta \vec{u}_p + \vec{U}.
\end{gather}

The $\mathcal{N}$ stands for the normal distribution.
It is noted that in the sampling process, the total energy conservation should be satisfied for the newly sampled stochastic particles in each cell.
As a result, the $\lambda_k$ carried by the stochastic particle is determined.
At the beginning of each time step, the collisionless particles from the wave component with macroscopic variables $\rho^* = e^{{-\Delta t}/{\tau_n}} \rho^h$, $\vec{U}$, $T_{thermal}$, and $\Theta_t = 0$, will be sampled and evolved in the time evolution.
For the above case, after obtaining the discretized particle velocity $\vec{u}_p$ based on Eq.\eqref{veldel-sampling} and Eq.\eqref{vel-sampling}, the $\lambda_k$ carried by the stochastic particles should satisfy the following relation
\begin{align*}
\frac{1}{2}\rho^* \vec{U}^2 + \frac{K+3}{2} \rho^* R T_{thermal}
= \sum \frac{m_p}{V} \frac{1}{2} \vec{u}_p^2
+ \sum \frac{m_p}{V} \frac{K+3}{2} \frac{1}{2\lambda_k},
\end{align*}
where $\lambda_k$ may differ from
$\lambda_{thermal} \overset{def}{=} 1/(2R T_{thermal})$, and $m_p$ can be directly determined from $\rho^*$ based on the mass conservation.


\setcounter{equation}{0}
\renewcommand\theequation{D.\arabic{equation}}
\section*{Appendix D: Results by WPTS in multiple simulations}
In addition to the previous presented results, another two results by WPTS are presented here. These two results are overall the best and worst cases we got in multiple simulations, which are shown in Figure \ref{fig:appendix-Fig-Rij-wp-ag4} and Figure \ref{fig:appendix-Fig-Rij-wp-ag5}, respectively.
In the best case, as shown in Figure \ref{fig:appendix-Fig-Rij-wp-ag4}, both the averaged Reynolds stress and instantaneous ones agree well with the reference solutions.
In the worst case shown in Figure \ref{fig:appendix-Fig-Rij-wp-ag5}, even though the results at $t_N = 1300$ show a larger difference, the averaged value is still satisfactory.
Overall, WPTS is capable of predicting the Reynolds stress well in the fully-developed CML study.
\begin{figure}[htbp]
	\centering
	\subfigure{
		\includegraphics[height=4.2cm]{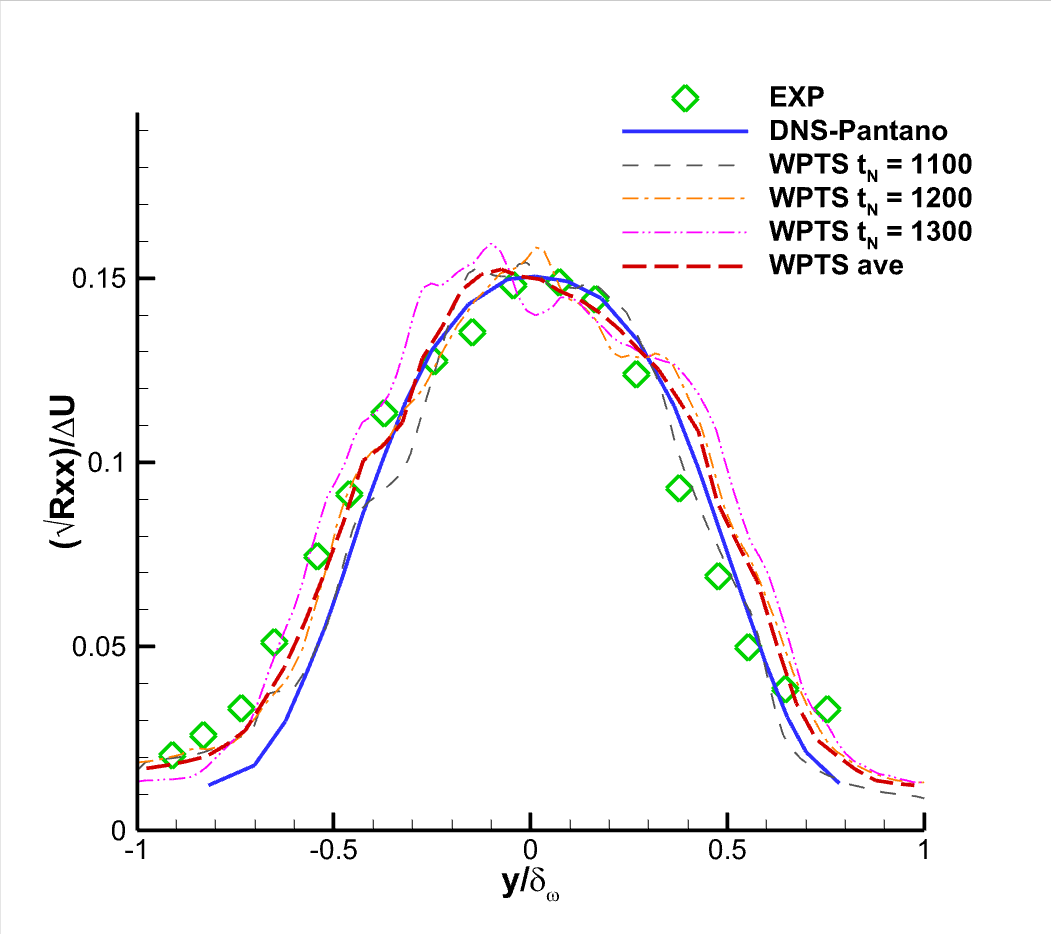}
	}
	\quad
	\subfigure{
		\includegraphics[height=4.2cm]{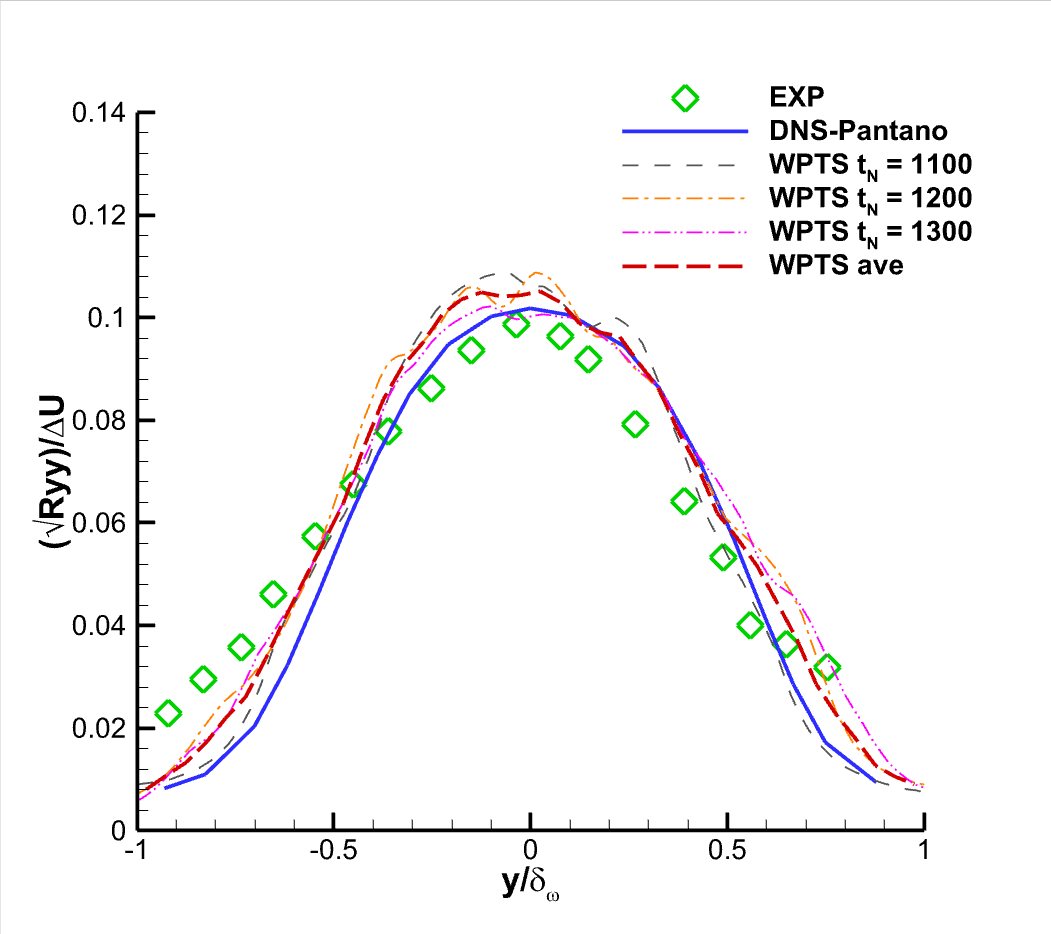}
	}
	\quad
	\subfigure{
		\includegraphics[height=4.2cm]{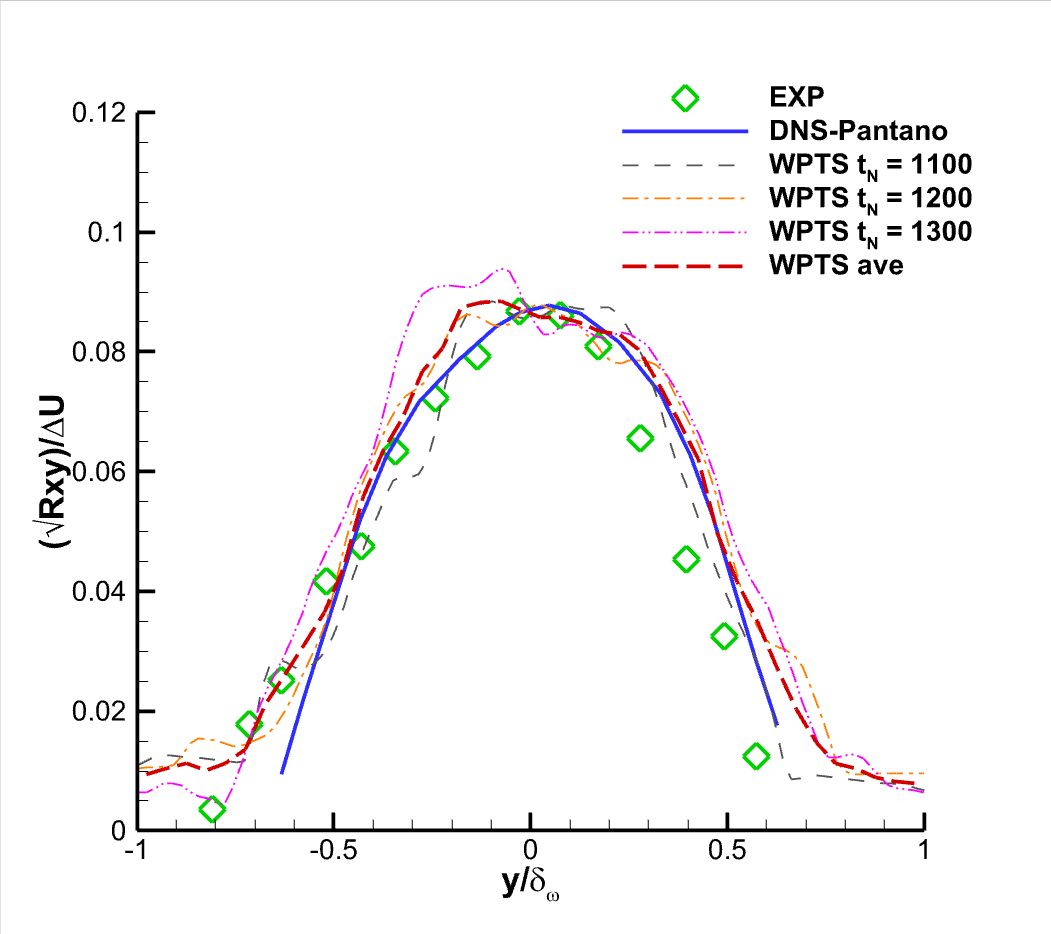}
	}
	\caption{The Reynolds stress terms $\sqrt{R_{xx}}/\Delta U$, $\sqrt{R_{yy}}/\Delta U$ and $\sqrt{R_{xy}}/\Delta U$ by WPTS at different times $t_{N} = 1100, 1200, 1300,$ and the averaged counterpart. The referent results by DNS and experiments are from \cite{Tur-case-mixing-DNS-pantano2002study} and \cite{Tur-case-mixing-EXP-elliott1990compressibility}, respectively.}
	\label{fig:appendix-Fig-Rij-wp-ag4}
\end{figure}

\begin{figure}[htbp]
	\centering
	\subfigure{
		\includegraphics[height=4.2cm]{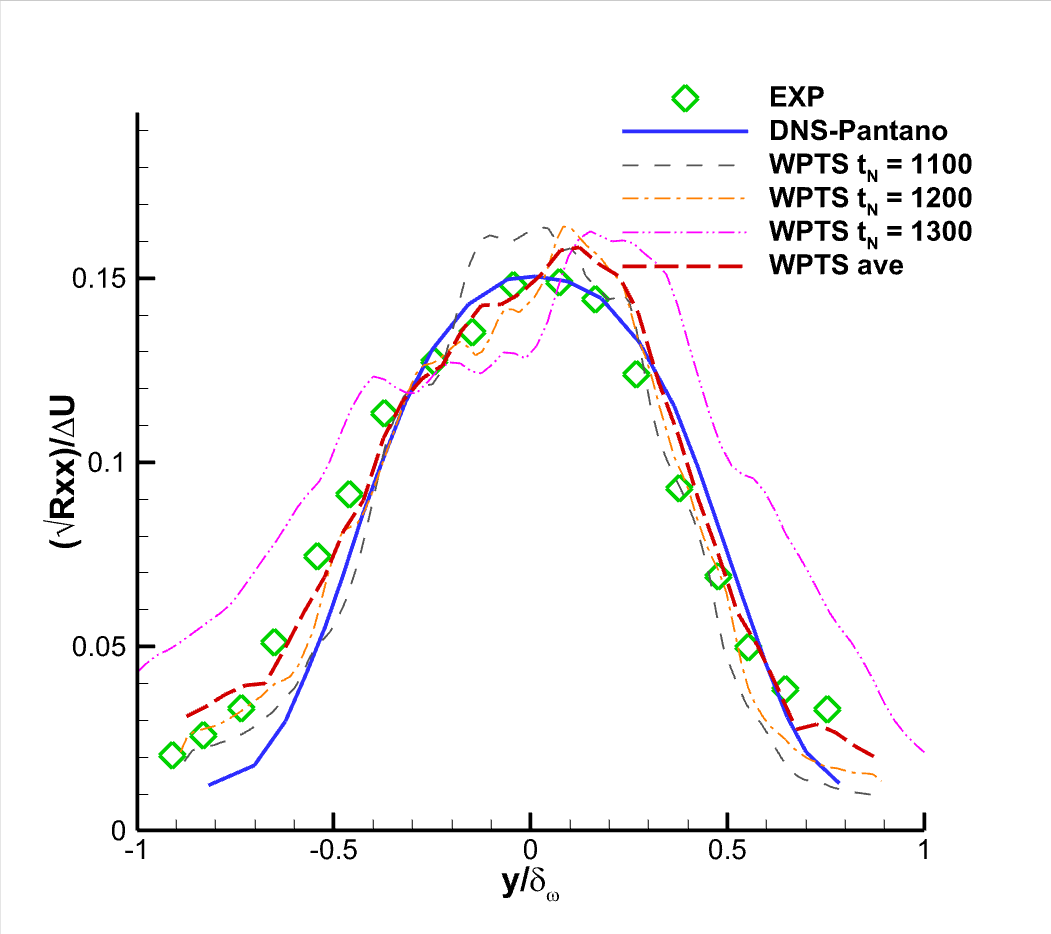}
	}
	\quad
	\subfigure{
		\includegraphics[height=4.2cm]{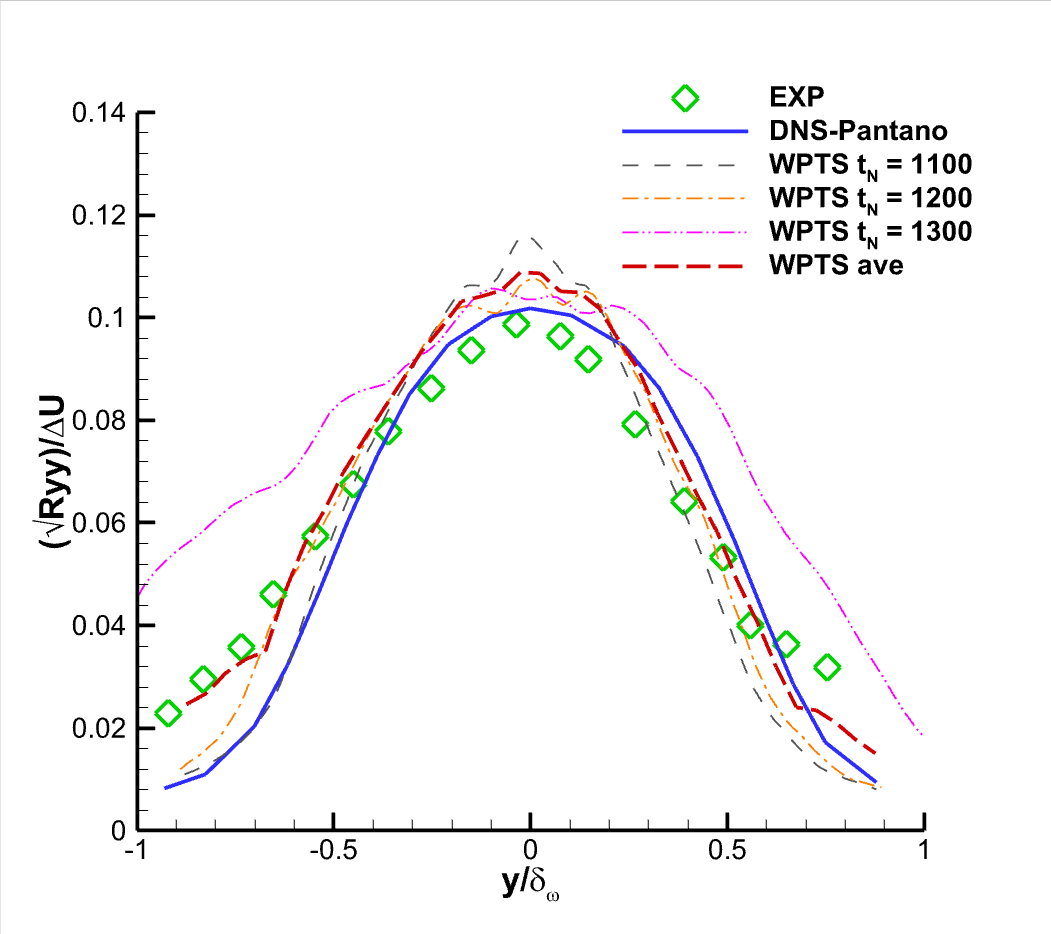}
	}
	\quad
	\subfigure{
		\includegraphics[height=4.2cm]{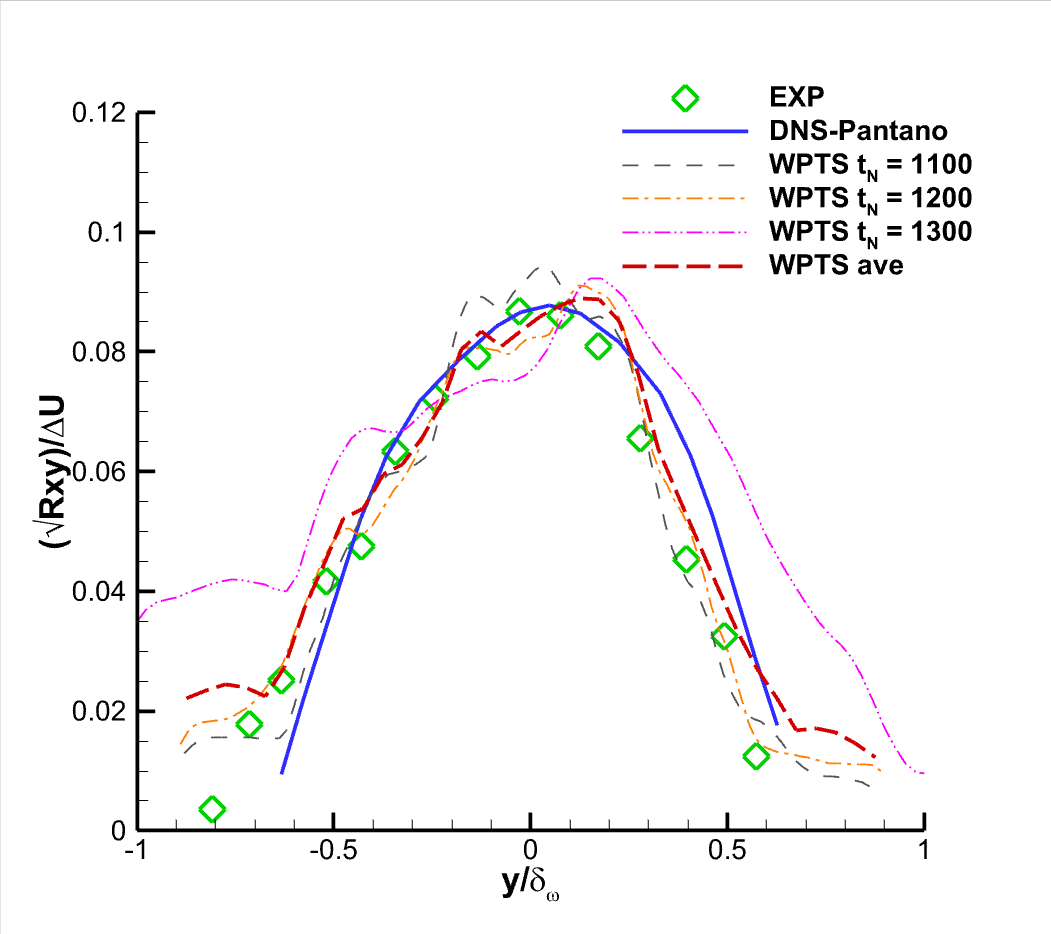}
	}
	\caption{The Reynolds stress terms $\sqrt{R_{xx}}/\Delta U$, $\sqrt{R_{yy}}/\Delta U$ and $\sqrt{R_{xy}}/\Delta U$ by WPTS at different times $t_{N} = 1100, 1200, 1300,$ and the averaged counterpart. The referent results by DNS and experiments are from \cite{Tur-case-mixing-DNS-pantano2002study} and \cite{Tur-case-mixing-EXP-elliott1990compressibility}, respectively.}
	\label{fig:appendix-Fig-Rij-wp-ag5}
\end{figure}

\bibliographystyle{plain}%
\bibliography{reference}
\end{document}